\documentclass[aps,prl,reprint,amsfonts,amssymb,amsmath,superscriptaddress,longbibliography,floatfix]{revtex4-2}

\pdfoutput=1

\usepackage{amssymb,amsmath,bm}
\usepackage{fullpage}
\usepackage{graphicx}
\usepackage{bibunits}
\usepackage{amsmath}
\usepackage{wrapfig}
\usepackage{floatflt}
\usepackage{color,soul}
\usepackage{xspace}
\usepackage{dsfont}
\usepackage[official,right]{eurosym}
\usepackage[top=0.8in, bottom=0.8in, left=0.75in, right=0.75in]{geometry}
\usepackage[utf8]{inputenc}
\usepackage[colorlinks=true,urlcolor=blue,citecolor=blue,linkcolor=blue,bookmarks=false]{hyperref}
\usepackage[flushleft]{threeparttable}
\usepackage{array,booktabs,makecell}
\usepackage{tabularx,array}
\newcolumntype{Y}{>{\centering\arraybackslash}X}
\usepackage{csquotes}
\MakeOuterQuote{"}
\usepackage{cleveref}
\usepackage[T1]{fontenc}

\newcommand{\nbwo}{Nd$_{3}$BWO$_9$\xspace}

\newcommand{\pbwo}{Pr$_{3}$BWO$_9$\xspace}

\newcommand{\ket}[1]{\left| #1 \right\rangle}

\newcommand{\be}{\begin{equation}}
	\newcommand{\ee}{\end{equation} }
\newcommand{\bea}{\begin{eqnarray} }
	\newcommand{\eea}{\end{eqnarray} }

\begin{document}
	
	\title{Braided Ising spin-tube physics in a purported kagome magnet}
	\author{J.~Nagl}
	\email{jnagl@ethz.ch}
	\affiliation{Laboratory for Solid State Physics, ETH Z{\"u}rich, 8093 Z{\"u}rich, Switzerland}
	
	\author{D.~Flavi{\'a}n}
	\affiliation{Laboratory for Solid State Physics, ETH Z{\"u}rich, 8093 Z{\"u}rich, Switzerland}
	
	\author{B.~Duncan}
	\affiliation{Laboratory for Solid State Physics, ETH Z{\"u}rich, 8093 Z{\"u}rich, Switzerland}
	
	\author{S.~Hayashida}
	\affiliation{Neutron Science and Technology Center, Comprehensive Research Organization for Science and Society (CROSS), Tokai, Ibaraki 319-1106, Japan}
	
	\author{O.~Zaharko}
	\affiliation{Laboratory for Neutron Scattering and Imaging, PSI Center for Neutron and Muon Sciences, Forschungsstrasse 111, 5232 Villigen, PSI, Switzerland}
	
	\author{E.~Ressouche}
	\affiliation{Université Grenoble Alpes, CEA, IRIG, MEM, MDN, 38000 Grenoble, France}
	
	\author{J.~Ollivier}
	\affiliation{Institut Laue-Langevin, 71 Avenue des Martyrs, CS 20156, 38042 Grenoble Cedex 9, France}
	
	\author{Z.~Yan}
	\affiliation{Laboratory for Solid State Physics, ETH Z{\"u}rich, 8093 Z{\"u}rich, Switzerland}
	
	\author{S.~Gvasaliya}
	\affiliation{Laboratory for Solid State Physics, ETH Z{\"u}rich, 8093 Z{\"u}rich, Switzerland}
	
	\author{A. Zheludev}
	\email{zhelud@ethz.ch}
	\homepage{http://www.neutron.ethz.ch/}
	\affiliation{Laboratory for Solid State Physics, ETH Z{\"u}rich, 8093 Z{\"u}rich, Switzerland}
	
	\date{\today}

	\begin{abstract}
		The magnetic insulator \nbwo has been previously proposed to realize the highly frustrated breathing kagome lattice model. We report a combination of single-crystal neutron scattering studies and numerical simulations that debunk this interpretation. We show that it is the {\it inter}-plane couplings that determine the physics. To explain the exotic magnetism, we derive a simple {\it one-dimensional} Ising model composed of twisted triangular spin-tubes, {\it i.e.}, triple braids of Ising spin chains with almost-orthogonal anisotropy frames and competing ferro-antiferromagnetic interactions. This model can account for the ground state, excitations, the numerous field-induced fractional magnetization plateau phases and incommensurate magnetic correlations at elevated temperatures. \nbwo constitutes a rare example where rich magnetic phenomena can be understood and simulated quantitatively in terms of a simple classical Hamiltonian.
	\end{abstract}
	
	\maketitle
	
	Spin systems with strong geometric frustration are the most likely to host exotic quantum magnetic phases \cite{diepFrustratedSpinSystems2020}. One of the oldest known and most important models of this type is an antiferromagnet (AFM) on the so-called kagome lattice. It can be viewed as a partially depleted triangular lattice, but has a much higher degree of frustration.  Precious few experimental realizations of kagome AFMs have been found to date. Unfortunately, their physics is complicated by either additional "undesirable" terms in the Hamiltonian \cite{zorkoCoexistenceMagneticOrder2019, chatterjeeSpinLiquidMagnetic2023} or intrinsic structural disorder \cite{hanCorrelatedImpuritiesIntrinsic2016}. This is one reason why the recent discovery of an entirely new family of rare-earth compounds R$_3$BWO$_9$ \cite{ashtarNewFamilyDisorderFree2020} featuring disorder-free "breathing-kagome" \cite{schafferQuantumSpinLiquid2017a} arrangements of magnetic ions has caused a great deal of excitement and intense experimental investigation. As hoped for, the latter revealed exotic frustration phenomena and complex magnetic phase diagrams  \cite{flavianMagneticPhaseDiagram2023a, songMagneticFieldTuned2023, yadavMagneticPropertiesFieldinduced2024, naglExcitationSpectrumSpin2024a, zengLocalEvidenceCollective2021, zengIncommensurateMagneticOrder2022}.
	
	In this Letter we report single-crystal neutron scattering, magnetic torque and dilatometry studies, as well as  numerical simulations for one particular species, namely \nbwo \cite{flavianMagneticPhaseDiagram2023a, songMagneticFieldTuned2023, yadavMagneticPropertiesFieldinduced2024}. We conclude that, due to peculiarities of single-ion electronic configurations, the breathing-kagome spin arrangement is largely irrelevant. The highly peculiar magnetic properties, including a plethora of magnetization-plateau phases and a commensurate-to-incommensurate transition, are due to an entirely different mechanism of frustration. The system is best described as braids of Ising spin chains with misaligned and almost-orthogonal anisotropy frames for individual spins. 
	%This model is able to {\em quantitatively} account for the observed magnetic ground states, phase transitions, excitations, finite-$T$ thermodynamics and correlations. 
		
\begin{figure}[tbp]
		\includegraphics[scale=1]{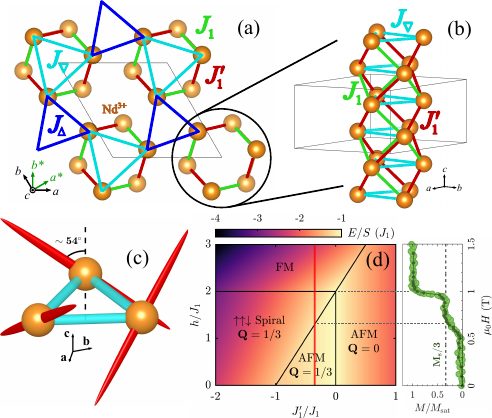}
		\caption{(a) Exchange pathways in \nbwo, emphasizing the breathing kagome structure realized in the $\mathbf{ab}$-plane. (b) The shortest bonds connect Nd$^{3+}$ ions in adjacent kagome planes through inequivalent superexchange pathways ($J_1'$)$J_1$, forming triple braids of spin-chains - i.e. frustrated triangular spin-tubes - running (counter-)clockwise along the $\mathbf{c}$-axis. (c) Schematic of the Ising-like $g$-tensor ellipsoids for a triangle of spins, depicting the nearly orthogonal anisotropy frames. (d) Phase diagram of the single spin-tube model Eq.~\ref{eq:Hamilton} in an axial magnetic field. The exchange ratio $J_1'/J_1 \simeq -0.35$ (red line) can be estimated from the relative width of the $m_z = 1/3$ plateau in the magnetization data shown on the right \cite{flavianMagneticPhaseDiagram2023a}.}
		\label{fig:Crystal_structure}
	\end{figure}

	Like other members of the series, \nbwo crystallizes in hexagonal $P6_3$ structure. From the schematic shown in  Fig.~\ref{fig:Crystal_structure}a it is easy to see why it has been discussed in the context of breathing-kagome physics. The magnetic Nd$^{3+}$ ions indeed realize this type of lattice in the $\mathbf{ab}$-plane of the crystal. The two in-plane nearest-neighbor Nd-Nd distances are 4.25~\AA~and 4.90~\AA, respectively.  In Fig.~\ref{fig:Crystal_structure} the respective exchange constants are labeled  $J_\Delta$ and $J_\nabla$. Consistent with the resulting high degree of frustration, \nbwo shows a significant ratio of Weiss- to Néel temperature $|\theta_{\rm CW}|/T_N \gtrsim 10$, with AFM order observed only below $T_N \approx 0.3$ K. This picture is, however, incomplete. The shortest Nd-Nd bonds, 3.94~\AA, ($J_1$ and $J_1'$ in Figs.~\ref{fig:Crystal_structure}a,b with inequivalent Nd-O-Nd superexchange pathways) actually connect magnetic ions from {\em adjacent} planes \cite{flavianMagneticPhaseDiagram2023a}. Each of these bonds forms bundles of three intertwined spin chains running along the $\mathbf{c}$-axis, but twisted in opposite directions (Fig.~\ref{fig:Crystal_structure}b). Below we shall demonstrate that it is this one-dimensional "twisted-spin-tube" arrangement rather than the breathing-kagome structure that plays a decisive role for the magnetism.
	
	To understand the character of magnetic interactions, we first analyze the electronic state of individual ions.
	In \nbwo the Nd$^{3+}$ ions are in $4f^3$ electronic configuration and have trivial point symmetry ($C_1$), being caged by eight O$^{2-}$ ions.
	Previously reported high-energy inelastic neutron scattering (INS) spectra \cite{flavianMagneticPhaseDiagram2023a} show the first crystal electric field (CEF) excitation at 15.6~meV. Thus, only the lowest-energy Kramers doublet is relevant to the low-temperature properties.
	The low symmetry makes a direct determination of the 27 independent crystal field parameters unfeasible. Instead, we derive an approximate single-ion Hamiltonian from a point-charge calculation \cite{hutchingsPointChargeCalculationsEnergy1964}. We model the crystal electric field produced by the neighboring anions, approximating them as idealized point charges. For rare earth ions this approach is known to work very well \cite{dunEffectivePointchargeAnalysis2021}. Assuming a charge $-2e$ for each O$^{2-}$-site already gives an agreement with observed excitation energies to within $\sim 20\%$. By fitting the magnitude of the charge for each of the three inequivalent sites we obtain a near perfect agreement with INS data \cite{supplement}. The thus-determined CEF Hamiltonian can adequately reproduce both the temperature dependence of the excitations as observed with neutron spectroscopy, and the single-crystal magnetometry data down to a temperature of 2~K without any additional fitting \cite{supplement}.
	
	The ground state Kramers doublet is composed almost exclusively of the maximal angular momentum states $\ket{j^z = \pm 9/2}$, where the local quantization axes $z$ form an angle $\theta \simeq 54.4^{\circ}$ with the crystallographic $\mathbf{c}$-axis. For adjacent ions, these local anisotropy frames are rotated around $\mathbf{c}$ according to sixfold symmetry. As a result, and that is a crucial point, the anisotropy axes of nearest-neighbor ions in the breathing-kagome plane are almost {\it orthogonal}, forming an angle $\phi \simeq 89^{\circ}$ relative to one another (see Fig.~\ref{fig:Crystal_structure}c). The $g$-tensor of the ground state doublet is extremely anisotropic. From our CEF model we estimate $g_{zz}/g_{\bot} \simeq 20$ in the local frame of each ion and $g_{zz} \approx 6.19$. The near-orthogonality of anisotropy axes results in an almost isotropic magnetic susceptibility for the bulk sample \cite{flavianMagneticPhaseDiagram2023a}.
	
	The single-ion ground state imposes strict constraints on the magnetic interactions. The latter can be written as a general bilinear form $\sim \hat{j}_i^\alpha K_{ij}^{\alpha \beta'} \hat{j}_j^{\beta'}$ of the interacting moments $\hat{\bm{j}}$ in the {\it local} anisotropy frames of the two coupled ions. Since the transverse angular momentum components $\hat{j}^\pm$ only change $j^z$ by unity, they have no matrix elements within the ground state doublet \cite{supplement}. As a result, exchange interactions are rendered {\it classical}, taking an Ising form $\sim \hat{j}_i^z K_{ij}^{zz'} \hat{j}_j^{z'}$ where $z$ and $z'$ refer to the {\em  local} principal anisotropy axes. Furthermore, if we disregard the possibility of asymmetric (Dzyaloshinskii-Moriya) interactions, $K^{zz'}$ must vanish by symmetry when the respective $z$-axes are orthogonal. The far-reaching consequence is that the {\em in-plane interactions $J_\Delta$ and $J_\nabla$ are suppressed} and can be neglected.
	
	What remains are {\em nearly-decoupled twisted Ising spin-tubes}. Representing every Kramers doublet with a $S_{\rm eff}=1/2$ pseudospin, we arrive at an effective spin Hamiltonian for each tube:
	\begin{eqnarray}
		\mathcal{H}_{\rm eff} &=& \sum_{i,j} \left( J_1 \hat{S}^{z}_{i,j} \hat{S}^{z}_{i+1,j} + J'_1 \hat{S}^{z}_{i,j} \hat{S}^{z}_{i+1,(j-1) {\rm mod} 3}\nonumber \right) \\
		&-& g_{zz} \mu_0 \mu_{\rm B} \sum_{i,j} \mathbf{H} \cdot \mathbf{\Hat{z}} \, \hat{S}^z_{i,j}. \label{eq:Hamilton}
	\end{eqnarray}
	Here the index $i$ labels consecutive spins in each chain and $j=\{0,1,2\}$ refers to the three chains. In all cases, $\hat{\mathbf{z}}=\hat{\mathbf{z}}_{i,j}$ are unit vectors along the {\em local} anisotropy axes and $g_{zz}$ refers to the principal component of the $g$-tensor.
	
\begin{figure}[tbp]
		\includegraphics[scale=1]{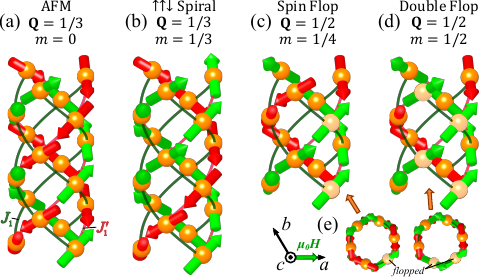}
		\caption{Magnetic structures realized for our single spin-tube model, including (a) the AFM state (zero field), (b) the $\uparrow \uparrow \downarrow$ spiral phase ($\mathbf{H} \parallel \mathbf{c}$) and (c,d) the spin-flop (SF) and double spin-flop (DSF) phases ($\mathbf{H} \bot \mathbf{c}$). (e) Top-down view of the latter two structures.}
		\label{fig:Magnetic_Structures}
	\end{figure}

	First, we consider the ground state of this tube-model. We evaluate the energies of all possible spin configurations, assuming a unit cell periodicity of $\leq 4$ along the $\mathbf{c}$-axis (up to 24 spins per tube). The resulting phase diagram for a field applied along the $\mathbf{c}$-axis is depicted in Fig.~\ref{fig:Crystal_structure}d.
	In zero field one finds three distinct phases: For $J_1, J_1' > 0$ the model is unfrustrated, realizing an AFM state with a 1-D propagation vector $q = 0$. Couplings of opposing sign $J_1'/J_1 < 0$ result in a tripling of the lattice period $q = 1/3$. A dominant FM $J_1'$ yields an $\uparrow \uparrow \downarrow$ spiral ground state with uniform magnetization $m_z = 1/3$ (Fig.~\ref{fig:Magnetic_Structures}b), whereas a dominant AFM $J_1$ results in a non-magnetized phase with AFM stacking along the $J_1$-bonds and $\uparrow \uparrow \uparrow \downarrow \downarrow \downarrow$-type correlations along $J_1'$ (Fig.~\ref{fig:Magnetic_Structures}a). The latter is the only state consistent with the experimentally observed $\mathbf{q} = (0,0,1/3)$ 3-D propagation vector in \nbwo and compensated magnetization in zero field \cite{flavianMagneticPhaseDiagram2023a}, and thus a strong contender for the magnetic ground state.
	
	To check this against experiment, we performed detailed single crystal neutron diffraction studies to determine the magnetic structure in zero field.  The data were collected on the lifting counter instruments D23 at ILL (France) \cite{d23data} and ZEBRA at PSI (Switzerland). Experimental details and group-theory based data analysis procedures are described in the Supplement \cite{supplement}. The obtained magnetic structure belongs to the $\Gamma_2$ representation of the symmetry group.
	This is consistent with the AFM $q=1/3$ phase predicted for our Ising tube model (see Fig.~\ref{fig:Magnetic_Structures}a).  Moreover, the refined orientations all magnetic moments in \nbwo is less than 4$^\circ$ away from the easy axes predicted by the point-charge model.
	
	We conclude that \nbwo is in the frustrated regime $J_1 > -J_1' > 0$. While each individual chain is clearly {\it unfrustrated}, these chains are braided together, intertwining at every third lattice position (non-bipartite) to give rise to a highly frustrated exchange topology.
	In an axial magnetic field $\mathbf{H \parallel c}$, the $\uparrow \uparrow \downarrow$ spiral phase is stabilized before the fully polarized ferromagnetic state, resulting in a fractional $m_z = 1/3$ magnetization plateau with $q = 1/3$, as seen experimentally. From the width of this plateau phase $h_{\rm c} / h_{\rm sat} = 1 + J_1'/J_1 \simeq 0.65$ (see Fig.~\ref{fig:Crystal_structure}(d)), we can estimate the exchange ratio $J_1'/J_1 \approx -0.35(5)$ for the spin-tube couplings. This is very close to the value $J_1' \approx -0.37 J_1$ determined in the sister compound \pbwo from INS data \cite{naglExcitationSpectrumSpin2024a}. Of course, our purely 1-D model is unable to account for the  stacking of adjacent tubes in the 3-D crystal, which is governed by weak residual inter-tube exchange or dipolar interactions \cite{supplement}.
	
	\begin{figure}[tbp]
		\includegraphics[scale=1]{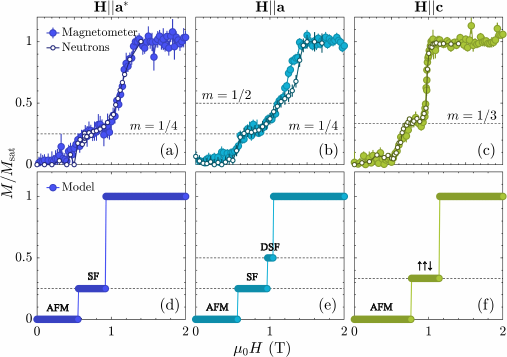}
		\caption{(a-c) Magnetization curves in \nbwo taken at $T \lesssim 120$ mK for various field directions using Faraday balance magnetometry (large symbols) and neutron diffraction (small symbols). The neutron data in (b) are from this work, the rest is from Ref.~\cite{flavianMagneticPhaseDiagram2023a}. (d-f) Model calculations of the same magnetization curves using $J_1 = 0.24$ meV and $J_1' = -0.35 J_1$, showing excellent qualitative agreement for all phases. In (e) we assume a $4^\circ$ misalignment of the field in the $\mathbf{ac}$-plane.}
		\label{fig:Magnetization}
	\end{figure}
	
	Additional plateau phases appear in magnetic fields applied in the plane. Fig.~\ref{fig:Magnetization}a-c summarizes magnetization curves measured using Faraday balance and neutron diffraction magnetometry (NDM) ~\cite{flavianMagneticPhaseDiagram2023a}. For $\mathbf{H}\|\mathbf{a}$ new NDM data were collected using the ZEBRA diffraction setup with a 1.8 T horizontal cryomagnet, by tracking the field dependence of the $(002)$ structural reflection \cite{supplement}.  The prominent plateau seen in low transverse fields has been previously assigned fractional values of  $m_z = 1/4$ or $m_z = 1/3$ \cite{flavianMagneticPhaseDiagram2023a, songMagneticFieldTuned2023}. The new measurements and a careful re-examination of data of Ref.~\cite{flavianMagneticPhaseDiagram2023a} clearly favor $m_z = 1/4$. An additional weak kink can be seen around $\mu_0 H \sim 1.2$ T, pointing to yet another plateau, with $m_z = 1/2$. It is barely visible in previous  Faraday-balance experiments~\cite{flavianMagneticPhaseDiagram2023a}, but quite prominent in the new NDM dataset. That it corresponds to a distinct thermodynamic phase is separately confirmed by magnetic torque and dilatometry measurements \cite{supplement}. While the presence of a multitude of plateau states is not unheard of in highly frustrated 2-D geometries such as Shustry-Sutherland \cite{matsudaMagnetizationSrCu2BO32013, qureshiPossibleStripePhases2022} or kagome \cite{nishimotoControllingFrustratedLiquids2013, gomezalbarracinSpinphononInducedMagnetic2013, okumaSeriesMagnonCrystals2019}, there have been no  1-D examples reported to date.
	
	\begin{figure*}[tbp]
		\includegraphics[scale=1]{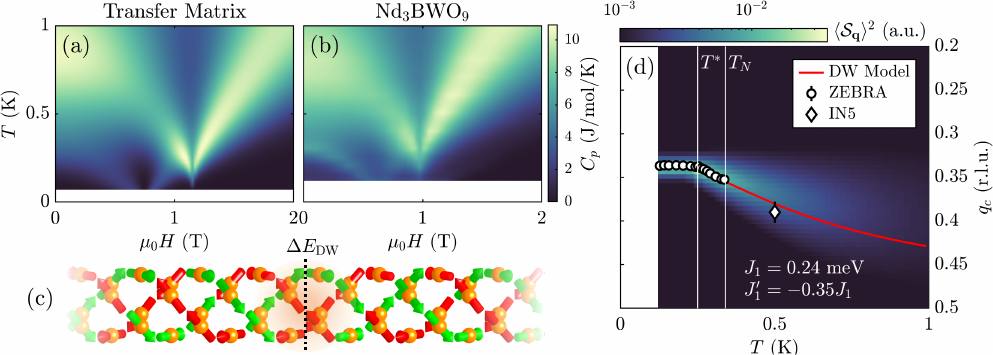}
		\caption{(a) Numerical transfer matrix calculation of the magnetic specific heat $C_m$ in an axial field $\mathbf{H \parallel c}$, visualized as a false color plot. (b) Direct comparison to experimental heat capacity in \nbwo for the same configuration (data taken from \cite{flavianMagneticPhaseDiagram2023a}). (c) Schematic of a domain wall defect. Notice how the $\uparrow \uparrow \uparrow$/$\downarrow \downarrow \downarrow$ pattern on every chain gives way to $\uparrow \uparrow \uparrow \uparrow$ and $\downarrow \downarrow \downarrow \downarrow$ clusters across the domain wall. (d) The $\mathbf{c}$-axis component of the magnetic propagation vector against temperature. Symbols: data from neutron scattering (circles: \cite{flavianMagneticPhaseDiagram2023a}, diamond: this work). Red line: the Boltzmann domain wall model, as described in the text. False color plot: spin correlation function from a classical Monte Carlo simulation.}
		\label{fig:Excitations}
	\end{figure*}
	
	All these plateau phases naturally occur in our Ising spin-tube model. Due to the canted anisotropy frames, all six Nd$^{3+}$ ions in the unit cell have different $g$-factors $g_{\rm eff} \approx g_{zz} \mathbf{\hat{H}} \cdot \mathbf{\hat{z}}_i$. As a result, a magnetic field applied in a general direction may influence some sites more than others, favoring different spin arrangements. For example, a $m_z = 1/4$ state with $q = 1/2$ is stabilized for all planar orientations $\mathbf{H \parallel ab}$, shown in Fig.~\ref{fig:Magnetic_Structures}c. It corresponds to a spin-flop (SF) phase where one spin with the largest $g$-factor is oriented along the field, while all others are stacked antiferromagnetically. A $m_z = 1/2$ state with $q = 1/2$ may occur at higher fields but is more fragile. For a strictly in-plane field it is only stable for a narrow range of orientations centered around $\phi \simeq 6.5^\circ$ from $\mathbf{H} \parallel \mathbf{a}$ (modulo 60$^\circ$) \cite{supplement}. This special angle corresponds to a high symmetry configuration where four ions share the same effective $g$-factor and the remaining two are fully decoupled from the field. Two spins become aligned with the field in this "double spin-flop" (DSF) phase, while the others remain staggered, as shown in Fig.~\ref{fig:Magnetic_Structures}d. Although this phase is not realized for $\mathbf{H} \parallel \mathbf{a}$ exactly, it is stabilized by a small out-of-plane field-component, appearing already for a tiny $\sim 2^\circ$ misalignment in the $\mathbf{ac}$-plane. This is well within the precision of aligning the sample in a cryomagnet and explains why this phase is repeatedly seen experimentally.

	Not only the presence of all these phases, but also the corresponding transition fields are quantitatively reproduced by our model. The computed magnetization curves are plotted in Fig.~\ref{fig:Magnetization}d--f for a direct comparison with experiment. Here we keep $J_1'/J_1 = -0.35$ fixed and use $J_1 = 0.24$ meV as precisely determined through neutron spectroscopy (see below). In order to reproduce the observed $m=1/2$ plateau, we assumed a $4^\circ$ misalignment of the field with the $\mathbf{a}$-axis in the $\mathbf{ac}$ plane. 
	
	A purely one-dimensional model with short range interactions is unable to support long-range order   at $T>0$.
	And yet, the finite-temperature thermodynamics of \nbwo is completely dominated by spin-tube physics. Specific heat computed in transfer-matrix simulations for a {\em single spin-tube} is shown in  Fig.~\ref{fig:Excitations}(a) as false-color plot against axial field and temperature \cite{supplement}. For comparison, experimental results from Ref.~\cite{flavianMagneticPhaseDiagram2023a} are shown in Fig.~\ref{fig:Excitations}(b). All the intricate features of the experiment are  reproduced. The only differences are the sharper $\lambda$-anomaly at $T_{\rm N}$ in the measurement and the small shift in the transition fields. Both discrepancies are due to inter-chain interactions and three-dimensional long range order in the real material. A similar match is achieved for the magnetic entropy \cite{supplement}. The strength of inter-chain interactions $J_{\rm eff}$ can be estimated by treating them at the mean-field (MF) level \cite{supplement}. Monte Carlo simulations of Eq.~\ref{eq:Hamilton} are used to compute the single-tube susceptibility at the ordering wave vector at a temperature equal to $T_{\rm N}$ of the real material. Solving a Weiss-style self-consistent MF equation we get $J_{\rm eff} \sim 0.2$~K, about $7 \%$ of the leading spin-tube exchange, in support of the weakly-coupled-spin-tube model for \nbwo.
	
	Next we turn to the low-energy spin excitations in \nbwo, measured on the IN5 cold-neutron disk-chopper spectrometer at ILL \cite{ollivierIN5ColdNeutron2011, in5data}. Probing these with INS is extremely challenging, because transitions between the $\ket{j^z = \pm 9/2}$ states expected to dominate this regime are forbidden by selection rules. Only the small $\sim 4\%$ admixing of lower angular momentum states into the single-ion ground state wavefunctions contributes to a finite scattering intensity. The spectrum is composed of two flat modes at $E_1 = 0.240(2)$~meV and $E_2 = 0.306(2)$~meV, respectively (see \cite{supplement}). This is consistent with expectations for an Ising magnet, where the dynamics is entirely local and due to flips of single spins. In our spin-tube model these modes correspond to flipping a spin at the edge (in the center) of an $\uparrow \uparrow \uparrow$ or $\downarrow \downarrow \downarrow$ spin cluster, costing an energy $E_1 = J_1$ ($E_2 = J_1 - J_1'$) respectively. This provides a direct measure of the dominant exchange parameters $J_1 \approx 0.240(2)$ meV and $J_1' \approx -0.28(2) J_1$, in good agreement with our previous estimate.
	
	One of the lingering puzzles about \nbwo was the incommensurate magnetic structure observed in zero field  at elevated temperatures \cite{flavianMagneticPhaseDiagram2023a}. The previously measured $\mathbf{c}$-axis component of the magnetic propagation vector  is plotted vs. temperature in circles in Fig.~\ref{fig:Excitations}d, showing a commensurate-lock-in  transition at $T^\ast \simeq 0.25$ K. As revealed by new neutron diffraction scans at even higher temperatures \cite{supplement}, in the paramagnetic state, the incommensurate Bragg reflections give way to a broad diffuse scattering peak. It too is incommensurate, the position being indicated by a diamond symbol in Fig.~\ref{fig:Excitations}d.
	
	For a quasi-1D system it is tempting to interpret incommensurate structures as signs of a quantum longitudinal spin density wave state \cite{kimuraLongitudinalSpinDensity2008, facherisSpinDensityWave2022, wuTomonagaLuttingerLiquid2019}. Our classical Ising spin-tube model provides a much simpler explanation. In a 1-D Ising magnet the low-energy dynamics is dominated by domain walls - kinks in the magnetic structure that act as phase slips in the local order parameter. These kinks are largely suppressed as $T \rightarrow 0$, but start to proliferate at finite temperatures, effectively changing the average periodicity of the magnetic structure and leading to incommensurate correlations.
	This is analogous to the modulated behavior seen in the axial next-nearest-neighbor Ising model, where entropically promoted soliton defects can form an extended incommensurate regime at elevated temperatures \cite{selkeANNNIModelTheoretical1988, bakIsingModelSolitons1980a}. In $d \geq 3$ dimensions, these solitons may even crystallize to form a "Devil's staircase" - a cascade of infinitely many commensurate phases.
	
	For our model, the lowest-energy kink is depicted in Fig.~\ref{fig:Excitations}c and has an energy cost of $\Delta E = (J_1 + J_1')/2 \approx 0.9$~K. It corresponds to a phase slip of $\pi/3$ in the magnetic structure. This allows for six different types of kink-bounded domains. Towards low temperatures (i.e. the dilute limit) we can ignore higher-lying excitations and attempt to calculate the change in the averaged lattice period by assuming Boltzmann statistics for the number of kinks $\langle n \rangle$. The resulting periodicity of magnetic correlations $\langle q \rangle = \frac{1+\langle n \rangle/N}{3}$ \cite{supplement} is plotted as a red line in Fig.~\ref{fig:Excitations}d. There is a clear crossover from commensurate to incommensurate behavior at $T^\ast$, in excellent agreement with experiment.
	As a cross check, we can reproduce this effect by performing classical Monte Carlo (MC) simulations. We model a single spin-tube of $L = 192$ unit cells (i.e. $\sim 10^4$ spins) with a simulated annealing procedure, keeping $J_1 = 0.24$ meV and $J_1' = -0.35 J_1$. In Fig.~\ref{fig:Excitations}d we show the calculated magnetic structure factor $\langle \mathcal{S}_\mathbf{q} \rangle^2$ as a false colorplot. Just as for the Boltzmann-domain wall model, the crossover at $T^*$ to incommensurate behavior is reproduced with no free parameters and a quasi-linear increase in $q$ that closely matches experiment is observed at high temperatures.
	
	In conclusion, despite previous speculations \cite{ashtarNewFamilyDisorderFree2020, songMagneticFieldTuned2023, yadavMagneticPropertiesFieldinduced2024}, \nbwo is hardly a model compound for kagome-lattice magnetism. It is an essentially classical quasi-1D Ising magnet with a peculiar frustration geometry of bundles of braided spin chains and an accidental cancellation of kagome-plane interactions. This unique geometry makes for a rare case where extremely rich magnetic behavior  can be understood and simulated quantitatively to astounding accuracy.

	\begin{acknowledgments}
		This work is supported by a MINT grant of the Swiss National Science Foundation. AZ thanks Dr. M. Zhitomirsky (CEA Grenoble) and Dr. I. Zaliznyak (Brookhaven National Lab) for insightful discussions. We also thank Prof. Titus Neupert and Dr. Nikita Astrakhantsev (University of Zürich) for illuminating discussions. We acknowledge the beam time allocation at PSI (ZEBRA id: 20220898, 20240065) and ILL (D23 id: CRG-3011, IN5 id: 4-05-861).
	\end{acknowledgments}

	\bibliography{Nd3BWO9_bib}
	%\nocite{*}
	
\end{document}

\title{Supplemental Material for "Braided Ising spin-tube physics in a purported kagome magnet"}
	\author{J.~Nagl}
	\email{jnagl@ethz.ch}
	\affiliation{Laboratory for Solid State Physics, ETH Z{\"u}rich, 8093 Z{\"u}rich, Switzerland}
	
	\author{D.~Flavi{\'a}n}
	\affiliation{Laboratory for Solid State Physics, ETH Z{\"u}rich, 8093 Z{\"u}rich, Switzerland}
	
	\author{B.~Duncan}
	\affiliation{Laboratory for Solid State Physics, ETH Z{\"u}rich, 8093 Z{\"u}rich, Switzerland}
	
	\author{S.~Hayashida}
	\affiliation{Neutron Science and Technology Center, Comprehensive Research Organization for Science and Society (CROSS), Tokai, Ibaraki 319-1106, Japan}

	\author{O.~Zaharko}
	\affiliation{Laboratory for Neutron Scattering and Imaging, PSI Center for Neutron and Muon Sciences, Forschungsstrasse 111, 5232 Villigen, PSI, Switzerland}

	\author{E.~Ressouche}
	\affiliation{Université Grenoble Alpes, CEA, IRIG, MEM, MDN, 38000 Grenoble, France}
	
	\author{J.~Ollivier}
	\affiliation{Institut Laue-Langevin, 71 Avenue des Martyrs, CS 20156, 38042 Grenoble Cedex 9, France}
	
	\author{Z.~Yan}
	\affiliation{Laboratory for Solid State Physics, ETH Z{\"u}rich, 8093 Z{\"u}rich, Switzerland}
	
	\author{S.~Gvasaliya}
	\affiliation{Laboratory for Solid State Physics, ETH Z{\"u}rich, 8093 Z{\"u}rich, Switzerland}
	
	\author{A. Zheludev}
	\email{zhelud@ethz.ch}
	\homepage{http://www.neutron.ethz.ch/}
	\affiliation{Laboratory for Solid State Physics, ETH Z{\"u}rich, 8093 Z{\"u}rich, Switzerland}
	
	\date{\today}
	\maketitle
	%\appendix
	\tableofcontents

	\section{Crystal Electric Field Hamiltonian}
	
	The large separation of energy scales $\mathcal{H}_{SO} \gg \mathcal{H}_{CEF} \gg \mathcal{H}_{ex}$ in rare earths allows us to consider separately the single-ion physics from the problem of two-ion interactions. We use susceptibility, magnetization and high-energy neutron spectroscopy data collected at $k_B T \gtrsim J_{ex}$ to refine the single-ion Hamiltonian, composed of a crystal field- and Zeeman contribution as
	\begin{gather}
		\mathcal{H}_{\rm CEF+Z} = \sum_{n,m} B_n^m \mathcal{O}_n^m - \mu_B g_J \mathbf{H} \cdot \mathbf{J},
	\end{gather}
	where $\mathcal{O}_n^m$ are the Stevens operators and $g_J$ is the Landé $g$-factor. In \nbwo, each Nd$^{3+}$ ion is surrounded by eight O$^{2-}$ ligands, all placed at slightly different distances. Thus, there are no point symmetries at the magnetic site, allowing for 27 independent crystal field parameters $B_n^m$. This means any unrestricted fit of experimental data will be underconstrained and cannot be used to uniquely identify the Hamiltonian.
	
	Instead we follow the strategy in \cite{naglExcitationSpectrumSpin2024a} and rely on a point charge approximation \cite{hutchingsPointChargeCalculationsEnergy1964, dunEffectivePointchargeAnalysis2021}. An estimate of the crystal field parameters is obtained directly from the experimentally determined chemical structure by calculating the electrostatic repulsion of neighboring ligands. By diagonalizing this parameter-free Hamiltonian, we obtain the single-ion spectrum and wavefunctions, allowing a direct comparison to experiment. 
	This can be further improved by adopting an {\it effective} point charge model, i.e. allowing the charges of the three inequivalent O$^{2-}$ sites to vary from $q_i = -2e$ to account for imperfections in the structural refinement, finite extent of charge clouds, etc. As we will show below, this model can quantitatively reproduce our neutron scattering and magnetometry data, allowing for an accurate approximation of the single-ion Hamiltonian. Our CEF calculations are performed using the PyCrystalField software \cite{scheiePyCrystalFieldSoftwareCalculation2021}.
	
	\subsection{Fit Results}
	
\begin{figure}[tbp]
		\includegraphics[scale=1]{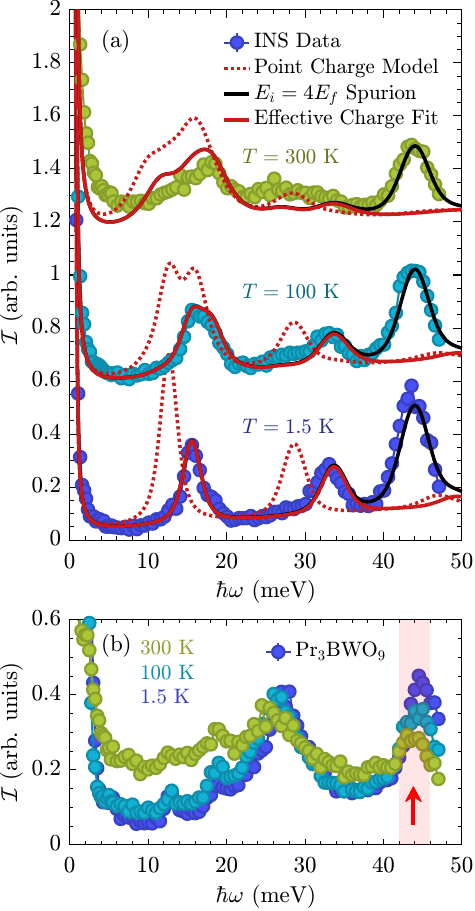}
		\caption{(a) INS powder spectra of \nbwo at various temperatures \cite{flavianMagneticPhaseDiagram2023a} compared to the parameter-free point charge model and the effective point charge fit. The 44 meV mode follows the intensity of the elastic line and is likely attributed to $E_i = 4E_f$ higher harmonic scattering. Elevated temperatures are offset by 0.6 units per curve for clarity. (b) The same "spurion"-mode is also observed in analogous INS spectra of the isostructural \pbwo species.}
		\label{fig:Supp_CEF}
	\end{figure}

	In Fig.~\ref{fig:Supp_CEF}(a) we present a direct comparison of the previously reported INS powder spectra \cite{flavianMagneticPhaseDiagram2023a} (symbols) to our point charge model calculations (lines). 
	
	First, we comment on the intense peak at 44~meV energy transfer, which has previously been assigned as a CEF excitation of Nd \cite{flavianMagneticPhaseDiagram2023a, yadavMagneticPropertiesFieldinduced2024}. We have strong evidence that in our data it is spurious, or at least substantially contaminated by spurious scattering. It corresponds to the scattering configuration where $E_i = 4 E_f$, and is due to elastic scattering in the sample and $\lambda/2$ scattering in the monochromator. These "spurions" are a known and well-document problem in thermal 3-axis instruments \cite{ShiraneNeutronTripleAxis2002}. Its observed temperature dependence is consistent with that of the incoherent elastic line (Debye-Waller factor). An almost identical peak was observed in the sister material \pbwo, as shown in Fig.~\ref{fig:Supp_CEF}(b). We have therefore excluded this feature from our analysis.
	
	In Fig.~\ref{fig:Supp_CEF} the calculations are based on the initial point charge model are shown in a dotted line. It reproduces qualitatively the observed excitation energies, but cannot give a quantitative agreement of intensities.
	The solid line is the  effective-charge model fit to the $T = 1.5$ K data.  The fit is fully constrained, with four fit parameters (three charges and an overall intensity prefactor) and as many experimental quantities (two CEF excitation energies and their respective intensities). The resulting CEF parameters are provided in Table~\ref{tab:CEF_par} for reference. The effective charges from the fit remain within 10\% of their nominal value $q = -2e$ (see Table~\ref{tab:CEF_fit}), indicating that only minor corrections are necessary to reproduce the data. The model reproduces the observed inelastic spectra rather well, including the high temperature data (no additional fitting). The small discrepancies seen at 300 K are due to additional phonon scattering.
		
	\begin{table}[tbp]
		\caption{CEF parameters of the effective point-charge model for the Nd$_1 = (0.084,0.724,0.353)$ site in the global $(a^{*},b,c)$ frame and the local frame of each ion, related to the global one through Euler angle rotations $(\alpha,\beta,\gamma) \approx (6.5^{\circ}, 54.4^{\circ}, 3.9^{\circ})$ in $z$-$y$-$z$ convention.}
		%\vspace{-0.5cm}
		\begin{tabularx}{0.45\textwidth}{YYY} \\
			\toprule\toprule
			$B_n^m$ ($10^{-2}$ meV) & Global Frame & Local Frame \\\midrule
			$B_2^{-2}$ & -0.812 & -8.489 \\
			$B_2^{-1}$ & -105.019 & -109.063 \\
			$B_2^0$ & 11.249 & 11.066 \\
			$B_2^1$ & -114.711 & -71.137 \\
			$B_2^2$ & -65.276 & -73.184 \\
			$B_4^{-4}$ & -1.362 & -1.366 \\
			$B_4^{-3}$ & 2.093 & 2.522 \\
			$B_4^{-2}$ & 0.301 & 0.229 \\
			$B_4^{-1}$ & 4.427 & 4.008 \\
			$B_4^0$ & 0.600 & 0.581 \\
			$B_4^1$ & -2.063 & -1.980 \\
			$B_4^2$ & -0.735 & -0.753 \\
			$B_4^3$ & -5.595 & -5.276 \\
			$B_4^4$ & -1.412 & -1.404 \\
			$B_6^{-6}$ & -0.072 & -0.062 \\
			$B_6^{-5}$ & -0.137 & -0.141 \\
			$B_6^{-4}$ & -0.064 & -0.062 \\
			$B_6^{-3}$ & 0.020 & 0.024 \\
			$B_6^{-2}$ & 0.036 & 0.033 \\
			$B_6^{-1}$ & -0.061 & -0.054 \\
			$B_6^0$ & -0.003 & -0.003 \\
			$B_6^1$ & -0.005 & -0.007 \\
			$B_6^2$ & 0.021 & 0.019 \\
			$B_6^3$ & -0.016 & -0.018 \\
			$B_6^4$ & -0.090 & -0.085 \\
			$B_6^5$ & -0.273 & -0.235 \\
			$B_6^6$ & -0.020 & -0.017 \\
			\toprule
		\end{tabularx}
		\label{tab:CEF_par}
	\end{table}

	\begin{table*}[tbp]
		\caption{The calculated CEF excitation energies, INS intensity ratios and effective ligand charges in the point charge model and effective charge fit are compared to experimental values in \nbwo.}
		\begin{tabularx}{0.8\linewidth}{ccYYYYcYcYYY}
			\midrule\midrule
			& & \multicolumn{4}{c}{CEF Energies (meV)} & & Intensity & & \multicolumn{3}{c}{Charges} \\\cline{3-6}\cline{8-8}\cline{10-12}
			\rule{0pt}{3ex}    
			& & $\Delta_{10}$ & $\Delta_{20}$ & $\Delta_{30}$ & $\Delta_{40}$ & & $\mathcal{I}_{10}/\mathcal{I}_{20}$ & & $q_1$ & $q_2$ & $q_3$ \\\midrule
			\nbwo & & 15.7 & 32.9 & - & - & & 1.30 & & -2 & -2 & -2 \\
			PCM & & 12.6 & 28.6 & 46.9 & 56.6 & & 1.88 & & -2 & -2 & -2 \\
			Effective PCM & & 15.6 & 33.7 & 49.2 & 60.7 & & 1.37 & & -2.18 & -2.01 & -2.12 \\
			\midrule\midrule
		\end{tabularx}
		\label{tab:CEF_fit}
	\end{table*}
	
	\begin{figure}[tbp]
		\includegraphics[scale=1]{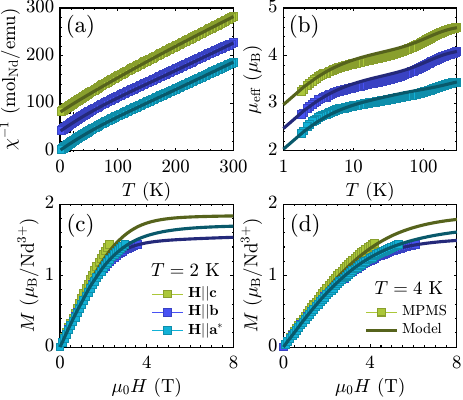}
		\caption{Magnetometry data on single crystals of \nbwo (squares) compared to model calculations based on the effective point charge model (lines), including a mean field exchange parameter $J_{\rm MF} = 0.22$ meV. We present the inverse magnetic susceptibility (a), effective moment $\mu_{\rm eff} \propto \sqrt{\chi T}$ (b) and magnetization curves at $T = 2$ K (c) and $T = 4$ K (d), collected along three principal crystallographic axes. An offset of 50 mol$_{\rm Nd}$/emu and 0.5 $\mu_{\rm B}$ per curve is added in (a) and (b) respectively for visibility.}
		\label{fig:Supp_Mag}
	\end{figure}
	
	Based on this model, we also calculate the magnetic susceptibility and magnetization of the system. As seen in Fig.~\ref{fig:Supp_Mag}, both exhibit excellent agreement with experiment for fields along all three principal crystallographic axes, including the small easy axis anisotropy seen in bulk. We note that a mean field exchange parameter $J_{\rm MF} \simeq 0.22$ meV is necessary to reproduce the magnetometry data, consistent with the small antiferromagnetic Weiss temperature. 	

	\subsection{Anisotropy Frames}
	
	Although our bulk magnetometry data seem to indicate nearly isotropic magnetism, the point charge calculations reveal that this does not hold for a single Nd$^{3+}$ ion. Instead, each magnetic moment is endowed with a strong easy-axis anisotropy with a {\it canted} local anisotropy frame. The nearly uniform susceptibility is only recovered upon averaging over the six ions in the unit cell, whose easy axes are {\it misaligned} but related by symmetry through $60^\circ$ rotations in the hexagonal plane.
	In Table~\ref{tab:WF} we compare the ground state wavefunctions $\ket{\pm}$ for the Nd$_1$ ion in the crystallographic $(a^{*},b,c)$ frame to those in the local anisotropy frame, related to the former by three Euler angle rotations $(\alpha,\beta,\gamma) \approx (6.5^{\circ},54.4^{\circ},3.9^{\circ})$ in $z$-$y$-$z$ convention. In the global frame the wavefunctions are complicated, involving non-zero complex weights for all $\ket{j^z}$ angular momentum states. But in the local frame of each ion, where the coordinate system is aligned with the anisotropy axes and the $g$-tensor becomes diagonal, the wavefunctions simplify considerably. The ground state doublet is nearly axially symmetric and over 95\% of the weights are concentrated on the maximal $\ket{j^z = \pm 9/2}$ components, indicating a nearly perfect Ising anisotropy. This is also reflected in the $g$-tensor, whose components can be evaluated directly from the ground state wavefunctions as
	\begin{align}
		g_{z \alpha} &= 2 g_J \bra{+} j^\alpha \ket{+} \\\nonumber
		g_{x \alpha} + i g_{y \alpha} &= 2 g_J \bra{-} j^\alpha \ket{+} \;\;\;\; \alpha \in \{x,y,z\}.
	\end{align}
	In the local frame, it reads
	\begin{gather}
		\mathbf{g} \approx 
		\begin{pmatrix}
			0.32 & 0 & 0 \\
			0 & 0.23 & 0 \\
			0 & 0 & 6.19
		\end{pmatrix}
	\end{gather}
	resulting in a highly anisotropic Zeeman splitting $g_{zz}/g_\perp \sim 20$.
		
	\begin{table}[tbp]
		\caption{Single-ion wavefunctions of the Nd$^{3+}$ ground state doublet obtained from effective point charge calculations in the global crystallographic frame $(a^*,b,c)$ [Nd$_1 = (0.084, 0.724, 0.353)$ lattice position] (a) and in the local frame of each ion (b), related to the global one given above through Euler angle rotations $(\alpha,\beta,\gamma) \approx (6.5^{\circ}, 54.4^{\circ}, 3.9^{\circ})$ in $z$-$y$-$z$ convention.}
		\begin{tabularx}{0.45\textwidth}{c|YY}
			\multicolumn{3}{c}{} \\
			\midrule\midrule
			(a) & $\ket{\psi_{+}}$ & $\ket{\psi_{-}}$ \\
			\midrule
			$\ket{-9/2}$ & \;$0.005+0.012i$ & $-0.351$ \\
			$\ket{-7/2}$ & \;$0.002+0.092i$ & \;\;\;$0.515+0.200i$ \\
			$\ket{-5/2}$ & \;$0.091+0.006i$ & $-0.471-0.197i$ \\
			$\ket{-3/2}$ & \;$0.133+0.052i$ & \;\;\;$0.401-0.064i$ \\
			$\ket{-1/2}$ & \;$0.203+0.056i$ & $-0.219+0.126i$ \\
			$\ket{+1/2}$ & \;$0.219+0.126i$ & \;\;\;$0.203-0.056i$ \\
			$\ket{+3/2}$ & \;$0.401+0.064i$ & $-0.133+0.052i$ \\
			$\ket{+5/2}$ & \;$0.471-0.197i$ & \;\;\;$0.091-0.006i$ \\
			$\ket{+7/2}$ & \;$0.515-0.200i$ & $-0.002+0.092i$ \\
			$\ket{+9/2}$ & \;$0.351$ & \;\;\;$0.005-0.012i$ \\
			\midrule\midrule
			
			(b) & $\ket{\psi_{+}}$ & $\ket{\psi_{-}}$ \\
			\midrule
			$\ket{-9/2}$ & $0$ & $-0.959$ \\
			$\ket{-7/2}$ & $-0.033-0.011i$ & \;\;\;$0.018-0.035i$ \\
			$\ket{-5/2}$ & \;\;\;$0.047+0.030i$ & $-0.094+0.144i$ \\
			$\ket{-3/2}$ & $-0.008-0.010i$ & \;\;\;$0.102+0.158i$ \\
			$\ket{-1/2}$ & \;\;\;$0.019+0.088i$ & $-0.013+0.034i$ \\
			$\ket{+1/2}$ & \;\;\;$0.013+0.034i$ &  \;\;\;$0.019-0.088i$ \\
			$\ket{+3/2}$ & \;\;\;$0.102-0.158i$ & \;\;\;$0.008-0.010i$ \\
			$\ket{+5/2}$ & \;\;\;$0.094+0.144i$ & \;\;\;$0.047-0.030i$ \\
			$\ket{+7/2}$ & \;\;\;$0.018+0.035i$ & \;\;\;$0.033-0.011i$ \\
			$\ket{+9/2}$ & $0.959$ & $0$ \\
			\midrule\midrule
		\end{tabularx}
		\label{tab:WF}
	\end{table}	

	\subsection{Ground State Projection}

	Since the crystal-field splitting $\Delta_{10} \sim 180$ K to the first excited state is very large, only the lowest Kramers doublet states are populated at low-$T$ where exchange interactions become relevant. The magnetic/thermodynamic properties are then fully determined by these low-energy degrees of freedom, allowing us to construct a simple effective Hamiltonian by projecting down to the lowest doublet $\ket{\pm} \simeq \ket{j^z = \pm 9/2}$. The local pseudospin operators $\mathbf{\hat{S}}_i$ in this restricted subspace can be written as
	\begin{gather}
		\hat{S}^z_i = \frac{1}{2} (\ket{+}_i\bra{+}_i - \ket{-}_i\bra{-}_i), \quad \hat{S}^\pm_i = \ket{\pm}_i\bra{\mp}_i.
	\end{gather}
	They are directly related to the magnetic dipole moment
	\begin{gather}
		\hat{\bm{\mu}}_i = -g_J \mu_{\rm B} \hat{\mathcal{P}} \hat{\bm{j}}_i \hat{\mathcal{P}} \simeq -\mu_{\rm B} g_{zz} \mathbf{\Hat{z}}_i \hat{S}^z_i
	\end{gather}
	where $\mathcal{\hat{P}} = \ket{+}\bra{+}-\ket{-}\bra{-}$ represents the ground state projector, $g_{zz} = \pm 2 g_J \bra{\pm} j^z \ket{\pm} \approx 6.19$ is the longitudinal $g$-factor and we dropped the $\times 20$ smaller transverse component. The ground state doublet is well approximated by the pure $\ket{\pm} \approx \ket{j^z = \pm 9/2}$ states, resulting in a nearly ideal Ising model Hamiltonian where only $\hat{S}^z$ contributes to the magnetic dipole moment.
	
	\subsection{Two-Ion Interactions}
	
	The dominant two-ion interactions in rare-earth magnetic insulators are the (super-)exchange and the magnetostatic dipolar coupling. As explained in the main text, it is sufficient to consider only the bare projection of the microscopic Nd$^{3+}$-Nd$^{3+}$ interactions into the ground state doublet, expressed in terms of our pseudospins. The vanishing matrix elements $\bra{\mp} \hat{j}^\pm \ket{\pm} \approx 0$ in the local frame directly result in an Ising-form for $any$ bilinear two-ion interaction, be it from superexchange or dipolar coupling. We note that higher multipolar interactions are allowed by symmetry, but are restricted to rank $\leq 7$ by the maximal amount of total angular momentum transferred by the $4f$ electron in each step of the superexchange process. Thus, they cannot connect the leading $\ket{j^z = \pm 9/2}$ ground state components and can be ignored \cite{iwaharaExchangeInteractionMultiplets2015, rauMagnitudeQuantumEffects2015}. 
	
	We stress that the main ingredient necessary to quench the in-plane interactions and render the remaining couplings Ising-like - a strong axial single-ion anisotropy with canting angle around $\theta \sim 55^\circ$ from the $c$-axis - can be deduced completely independently from the CEF analysis. The presence of a huge Ising anisotropy is evident from the low-energy spin excitations (Fig.~\ref{fig:IN5}), showing a dispersionless single spin-flip dynamics (and should be expected given the highly anisotropic NdO$_8$ cage with trivial $C_1$ point symmetry). Accepting this fact, a canting angle near $\theta \approx 56(3)^\circ$ (which directly determines the angle $\phi \approx 92(5)^\circ$ between breathing kagome moments) is the only way to recover the nearly uniform bulk magnetic susceptibility $\chi_c/\chi_{ab} \sim 1.08(5)$ observed at low temperatures \cite{flavianMagneticPhaseDiagram2023a}.
	
	In \nbwo, the strength of nearest neighbor dipole-dipole interactions amounts to $\mathcal{D} = \frac{\mu_0 (g_{\rm J} \mu_{\rm B} J)^2}{4 \pi r_{\rm nn}^3} \approx 0.1$ K, more than an order of magnitude weaker than the leading spin-tube coupling $J_1 \sim 3$ K seen in experiment. Therefore, dipolar coupling plays at most a sub-leading role and the two-ion interactions must be dominated by Nd-O-Nd superexchange processes. The associated bond angles and distances for the spin-tube and breathing kagome couplings are provided in table \ref{tab:bonds}. According to the Kanamori-Goodenough rules, all couplings should be antiferromagnetic except $J_1'$, with $\theta \sim 103^{\circ}$ close to the AF-F crossover. This may explain the opposing signs of $J_1$ and $J_1'$ despite the visually similar bond structure. Interestingly, the weaker dipolar coupling only reinforces this trend: After projecting onto the local anisotropy directions $\hat{\mathbf{z}}_i$, the magnetostatic contribution to the leading spin-tube interactions amounts to $\mathcal{D}_1 \approx 0.07$ K and $\mathcal{D}_1' \approx -0.19$ K respectively, strongly discriminating between clockwise and counterclockwise bonds.
	
	\begin{table}[tbp]
		\caption{Bond distances and angles in \nbwo associated with the leading spin-tube / breathing kagome superexchange pathways discussed in the main text.}
		\begin{tabularx}{0.9\linewidth}{ccYYcY}
			\midrule\midrule
			& & \multicolumn{2}{c}{Distance (\AA)} & & Angle ($^{\circ}$) \\\cline{3-4}\cline{6-6}
			& & $r_{\mathrm{Nd-Nd}}$ & $r_{\mathrm{Nd-O}}$ & & $\theta_{\mathrm{Nd-O-Nd}}$ \\\midrule
			$J_1$ & & 3.94 & 2.41 - 2.56 & & 105 - 115 \\
			$J_1'$ & & 3.94 & 2.32 - 2.61 & & 103 \\
			$J_\Delta$ & & 4.25 & 2.45 - 2.56 & & 116 \\
			%$J_3$ & & 4.38 & 2.45 - 2.48 & & 125 \\
			$J_\nabla$ & & 4.90 & 2.55 - 2.61 & & 143 \\
			\midrule\midrule
		\end{tabularx}
		\label{tab:bonds}
	\end{table}

	\section{Neutron Diffraction Experiments}
	
	High quality single-crystal neutron diffraction data were collected at D23 (ILL, France) and ZEBRA (PSI, Switzerland). The same fully $^{11}$B-substituted single crystal of $m = 17$~mg was used in both experiments. The sample was mounted on the cold finger of a dilution refrigerator at $T \lesssim 0.1$ K, with the $\mathbf{ab}$ and $\mathbf{ac}$ crystallographic planes in the horizontal scattering plane of the instruments, respectively.
	
	\subsection{Nuclear Structure Determination}
	
	\begin{figure*}[tbp]
		\includegraphics[scale=1]{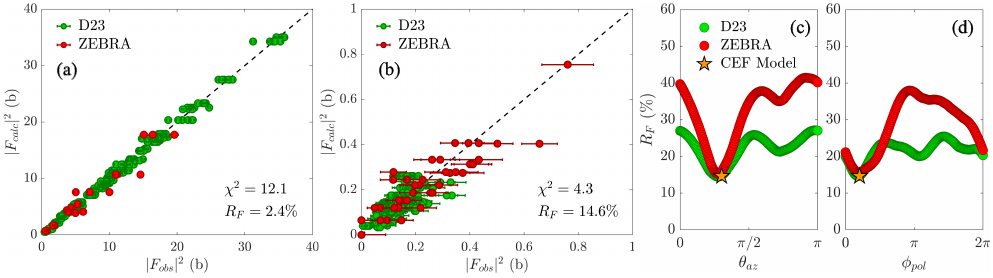}
		\caption{(a) Refinement of the observed nuclear Bragg intensities to the $P6_3$ crystal structure, considering two merohedral twins. (b) Refinement of the $\mathbf{q} = (0,0,1/3)$ magnetic intensities in zero field to the $\Gamma_2$ model discussed in the text. In each case the combined $R$-factor and goodness of fit from both diffraction experiments are quoted above. (c,d) Variation of the $R$-factor against azimuthal and polar angles in our magnetic structure model, confirming that the refined moment orientations are consistent with the local easy axes obtained from CEF measurements.}
		\label{fig:Supp_Refinement}
	\end{figure*}

	On D23, the nuclear structure was determined to high accuracy by collecting $\sim 320$ reflections at $T = 10$ K using a wavelength of $\lambda = 1.28$ \AA. The resulting integrated intensities were employed in a refinement with 22 fit parameters, i.e. 13 atomic positions, 6 isotropic Debye-Waller factors, 2 scale factors and 1 extinction coefficient. Due to the polar $P6_3$ structure with no preferred origin along $\mathbf{c}$, the $z$-component of the Boron atomic position was kept fixed at zero for the refinement. Furthermore, \nbwo is susceptible to a merohedral twinning \cite{chandraAnalysisCharacterizationData1999}, where each Bragg peak at $(h,k,l)$ contains also a contribution from the twinned $(k,h,-l)$ reflection. Both twins (i.e. 2 scale factors) need to be considered in order to obtain a reasonable fit. The refinement yields excellent agreement with experiment, with a final $R_F = 2.2$\% and $\chi^2 = 12.0$.
	An analogous procedure was carried out for the smaller ZEBRA dataset (25 reflections collected at $T = 1$ K using $\lambda = 2.3$ \AA), giving consistent results with $R_F = 7.2$\% and $\chi^2 = 13.1$. Both datasets show clear signs of merohedral twinning with a relative volume fraction of 2.0(1) : 1.	The combined nuclear refinement results are displayed in Fig.~\ref{fig:Supp_Refinement}(a), yielding a final $R_F = 2.4$\% and $\chi^2 = 12.1$.
	
	\subsection{Magnetic Structure Determination}
	
	For the magnetic structure, a total of $\sim 300$ reflections were collected at fractional $\mathbf{q} = (0,0,1/3)$ positions on D23 ($T = 50$ mK  and $\lambda = 1.28$ \AA) and ZEBRA ($T = 120$ mK and $\lambda = 1.38$ \AA). A symmetry analysis carried out in BasIreps \cite{rodriguez-carvajalRecentAdvancesMagnetic1993} restricts the space of possible spin configurations to six irreducible representations. Each is complex and appears three times in the decomposition, resulting in six fit parameters - these can be recast as the ordered moment, the polar- and azimuthal angles for one of the Nd ions and three complex phases. In Table~\ref{tab:SymmetryAnalysis} we provide a list of the general Fourier components for each of the six irreducible representations. Four of these irreps can be excluded based on systematic absences at $\mathbf{Q} = (0,0,2/3)$ in the corresponding magnetic space group formalism, inconsistent with experiment \cite{flavianMagneticPhaseDiagram2023a}. This leaves only the $\Gamma_2$ and $\Gamma_3$ representations (or $P6_5'$ and $P6_1'$ Shubnikov groups respectively). Based on CEF considerations (extremely strong axial single-ion anisotropy) we start by assuming a simplified model, setting the complex phases to zero and fixing the moment orientations along the local easy axes. With only one fit parameter, we find excellent agreement for $\Gamma_2$ with both D23 and ZEBRA datasets, resulting in a combined $R_F = 15$\% and $\chi^2 = 4.0$. This structure is consistent with the classical ground state of our Ising spin-tube model in the regime $J_1 > -J_1' > 0$, exactly the predicted spin configuration. All other irreps provide much worse agreement, with $R_F > 30$\%.
	
	\begin{figure}[tbp]
		\includegraphics[scale=1]{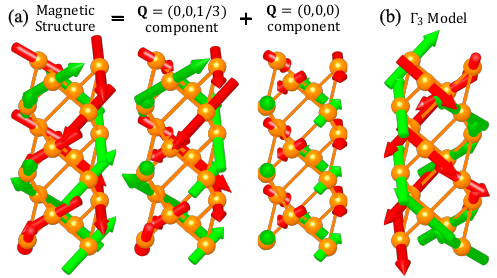}
		\caption{(a) The classical magnetic ground state predicted by our spin-tube model (left) can be decomposed into a modulated $\mathbf{q} = (0,0,1/3)$ component (center) and a uniform $\mathbf{q} = (0,0,0)$ contribution (right). Only the former can be probed accurately in experiment. (b) The alternative $\Gamma_3$ solution, corresponding to the same type of AFM ground state with opposite chirality.}
		\label{fig:Supp_MagStr}
	\end{figure}

	\begin{table*}[tbp]
		\caption{General Fourier coefficients for the six irreducible representations $\Gamma_{1-6}$ compatible with $\mathbf{q} = (0,0,1/3)$ propagation vector in \nbwo. Here we choose the $(0.084, 0.724, 0.353)$ lattice position as our Nd$_1$ site.}
		\renewcommand{\arraystretch}{1.2}
		\begin{tabularx}{1\linewidth}{ccYY|ccYY}
			\midrule\midrule
			IR & Atom & Symmetry & Fourier Coefficient $\mathbf{S}(q)$ & \hspace{0.2cm}IR\hspace{0.2cm} & Atom & Symmetry & Fourier Coefficient $\mathbf{S}(q)$ \\
			\midrule			
			$\Gamma_1$ & Nd$_1$ & $(x,y,z)$ & $(u,v,w)$ & 
			$\Gamma_4$ & Nd$_1$ & $(x,y,z)$ & $(u,v,w)$ \\
			& Nd$_2$ & $(-y,x-y,z)$ & $(-v,u-v,w)$ & 
			& Nd$_2$ & $(-y,x-y,z)$ & $(-v,u-v,w)$ \\
			& Nd$_3$ & $(-x+y,-x,z)$ & $(v-u,-u,w)$ & 
			& Nd$_3$ & $(-x+y,-x,z)$ & $(v-u,-u,w)$ \\
			& Nd$_4$ & $(-x,-y,z+\frac{1}{2})$ & $(-u,-v,w) \times e^{-i\pi/3}$ & 
			& Nd$_4$ & $(-x,-y,z+\frac{1}{2})$ & $(u,v,-w) \times e^{-i\pi/3}$ \\
			& Nd$_5$ & $(y,-x+y,z+\frac{1}{2})$ & $(v,v-u,w) \times e^{-i\pi/3}$ & 
			& Nd$_5$ & $(y,-x+y,z+\frac{1}{2})$ & $(-v,u-v,-w) \times e^{-i\pi/3}$ \\
			& Nd$_6$ & $(x-y,x,z+\frac{1}{2})$ & $(u-v,u,w) \times e^{-i\pi/3}$ & 
			& Nd$_6$ & $(x-y,x,z+\frac{1}{2})$ & $(v-u,-u,-w) \times e^{-i\pi/3}$ \\
			\midrule
			$\Gamma_2$ & Nd$_1$ & $(x,y,z)$ & $(u,v,w)$ & 
			$\Gamma_5$ & Nd$_1$ & $(x,y,z)$ & $(u,v,w)$ \\
			& Nd$_2$ & $(-y,x-y,z)$ & $(v,v-u,-w) \times e^{i\pi/3}$ & 
			& Nd$_2$ & $(-y,x-y,z)$ & $(v,v-u,-w) \times e^{i\pi/3}$ \\
			& Nd$_3$ & $(-x+y,-x,z)$ & $(u-v,u,-w) \times e^{-i\pi/3}$ & 
			& Nd$_3$ & $(-x+y,-x,z)$ & $(u-v,u,-w) \times e^{-i\pi/3}$ \\
			& Nd$_4$ & $(-x,-y,z+\frac{1}{2})$ & $(-u,-v,w) \times e^{-i\pi/3}$ & 
			& Nd$_4$ & $(-x,-y,z+\frac{1}{2})$ & $(u,v,-w) \times e^{-i\pi/3}$ \\
			& Nd$_5$ & $(y,-x+y,z+\frac{1}{2})$ & $(-v,u-v,-w)$ & 
			& Nd$_5$ & $(y,-x+y,z+\frac{1}{2})$ & $(v,v-u,w)$ \\
			& Nd$_6$ & $(x-y,x,z+\frac{1}{2})$ & $(u-v,u,w) \times e^{i\pi/3}$ & 
			& Nd$_6$ & $(x-y,x,z+\frac{1}{2})$ & $(v-u,-u,-w) \times e^{i\pi/3}$ \\
			\midrule
			$\Gamma_3$ & Nd$_1$ & $(x,y,z)$ & $(u,v,w)$ & 
			$\Gamma_6$ & Nd$_1$ & $(x,y,z)$ & $(u,v,w)$ \\
			& Nd$_2$ & $(-y,x-y,z)$ & $(v,v-u,-w) \times e^{-i\pi/3}$ & 
			& Nd$_2$ & $(-y,x-y,z)$ & $(v,v-u,-w) \times e^{-i\pi/3}$ \\
			& Nd$_3$ & $(-x+y,-x,z)$ & $(u-v,u,-w) \times e^{i\pi/3}$ & 
			& Nd$_3$ & $(-x+y,-x,z)$ & $(u-v,u,-w) \times e^{i\pi/3}$ \\
			& Nd$_4$ & $(-x,-y,z+\frac{1}{2})$ & $(-u,-v,w) \times e^{-i\pi/3}$ & 
			& Nd$_4$ & $(-x,-y,z+\frac{1}{2})$ & $(u,v,-w) \times e^{-i\pi/3}$ \\
			& Nd$_5$ & $(y,-x+y,z+\frac{1}{2})$ & $(v,v-u,w) \times e^{i\pi/3}$ & 
			& Nd$_5$ & $(y,-x+y,z+\frac{1}{2})$ & $(-v,u-v,-w) \times e^{i\pi/3}$ \\
			& Nd$_6$ & $(x-y,x,z+\frac{1}{2})$ & $(v-u,-u,-w)$ & 
			& Nd$_6$ & $(x-y,x,z+\frac{1}{2})$ & $(u-v,u,w)$ \\
			\midrule\midrule
		\end{tabularx}
		\label{tab:SymmetryAnalysis}
	\end{table*}
	
	We can refine also the moment orientations (polar and azimuthal angle), showing that the best-fit parameters in the $\Gamma_2$ model closely match the easy axes obtained from CEF measurements (see Fig.~\ref{fig:Supp_Refinement}(c,d)). Given this additional freedom, the $\Gamma_3$ irrep fits the data equally well with drastically different moment orientations. As seen in Fig.~\ref{fig:Supp_MagStr}(b), $\Gamma_3$ is an analogous solution to $\Gamma_2$ with opposite spin chirality: The winding direction of $J_1$ and $J_1'$ correlations is switched, and each spin is rotated by $\phi \rightarrow \phi + \pi$ in the basal plane. These two models have the same structure factor, but the $\Gamma_3$ solution, bearing completely different moment orientations, is inconsistent with the single-ion physics in this compound and can be excluded. As expected, the other four irreps still give much worse agreement.
	
	Finally, we can also vary the three complex phases in our refinement. These remain zero within uncertainties and do not improve the fit. Therefore, we conclude that the magnetic structure in zero field corresponds to the $\Gamma_2$ irrep or $P6_5'$ magnetic space group depicted in Fig.~\ref{fig:Supp_MagStr}(a), with each tube in the AFM $q=1/3$ phase. The combined refinement results using both D23 and ZEBRA datasets are shown in Fig.~\ref{fig:Supp_Refinement}(b), with a final $R_F = 14.6$\% and $\chi^2 = 4.3$. The corresponding fit parameters are provided in Table~\ref{tab:MagStr}, compared directly to the CEF model predictions. Both the ordered moment $m = 3.7(4)$ $\mu_B$ and the azimuthal and polar angles $\theta = 51(4)^\circ$ and $\phi = 35(10)^\circ$ are consistent with the CEF estimates within uncertainty.

	\begin{table*}[tbp]
		\caption{Magnetic structure parameters of \nbwo in zero field obtained from single crystal neutron diffraction, compared to the predicted values based on the CEF Hamiltonian.}
		
		\begin{tabularx}{1\linewidth}{YYYYYYY}
			\midrule\midrule
			& $\theta$ ($^\circ$) & $\phi$ ($^\circ$) & $\langle m \rangle$ ($\mu_B$) & $R_F$ (\%) & $\chi^2$ & $T$ (K) \\
			\midrule
			ZEBRA & 52(4) & 37(16) & 4.5(7) & 15.1 & 1.8 & 0.12 \\
			D23 & 48(8) & 34(12) & 3.5(3) & 14.2 & 3.8 & 0.05 \\
			CEF Model & 54.4 & 36.5 & 4.25 & - & - & 0 \\
			\midrule\midrule
		\end{tabularx}
		\label{tab:MagStr}
	\end{table*}            
		
	We note that the refined $\mathbf{q} = (0,0,1/3)$ component of the magnetic structure is a modulated spin density wave, consistent with the predicted classical ground state up to a small additional $\mathbf{q} = 0$ contribution to compensate the spin modulation (see Fig.~\ref{fig:Supp_MagStr}(a)). But the latter cannot reliably measured with unpolarized neutron diffraction because the weak magnetic peaks overlap with the much stronger nuclear reflections, leaving only the modulated components accessible to experiment.

	\subsection{Uniform Magnetization}
	
	Using the ZEBRA diffraction setup discussed above in conjunction with a 1.8 T horizontal cryomagnet, we investigated the field dependence of the (002) nuclear reflection for the $\mathbf{H \parallel a}$ configuration. In an external magnetic field, the induced ferromagnetic component to the magnetization density produces an extra scattering contribution at the nuclear positions, proportional to the square of the uniform magnetic moment. We selected the (002) reflection with a minimal nuclear scattering intensity and subtract this contribution, recovering the longitudinal magnetization as the square root of the field-dependent magnetic intensity (see main text). The $\mathbf{H \parallel a}$ field dependence of the $\mathbf{q} = (1/2,1/2,1/2)$ order parameter realized for the $m = 1/4$ and $m = 1/2$ plateau phases was determined in the same experiment (see below).
	
	\subsection{Diffraction Summary}
	
	\begin{figure*}[tbp]
		\includegraphics[scale=1]{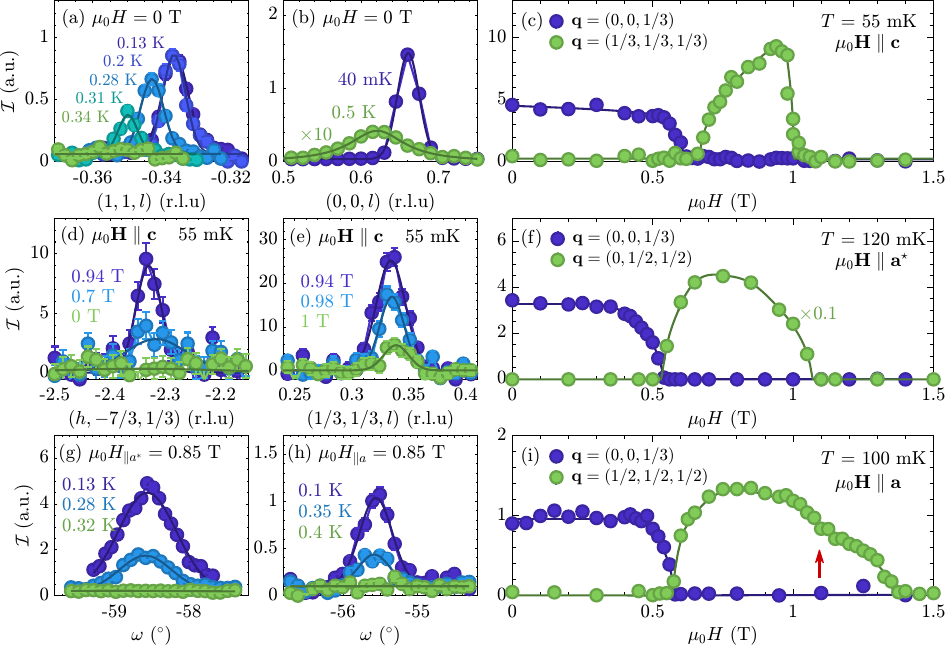}
		\caption{Diffraction scans in \nbwo at various fields and temperatures. (a) Temperature dependence of the $\mathbf{Q} \approx (1,1,-1/3)$ reflection, revealing a small incommensurability above $T^* \simeq 0.25$ K. (b) While the $\mathbf{Q} \approx (0,0,2/3)$ Bragg peak disappears at $T_{\rm N}$, a broad, incommensurate diffuse scattering maximum remains even at $T = 0.5$ K. (c) Field dependence of the magnetic propagation vectors $\mathbf{q}$ in $\mathbf{H \parallel c}$ configuration. (d-e) Exemplary diffraction scans of the $\mathbf{Q} = (-7/3,-7/3,1/3)$ and $\mathbf{Q} = (1/3,1/3,1/3)$ reflections in the $m = 1/3$ plateau phase. (f-i) Field dependence of the magnetic propagation vectors with in-plane field orientations $\mathbf{H \parallel a^*}$ and $\mathbf{H \parallel a}$. A temperature dependence of the $\mathbf{Q} = (0,1/2,1/2)$ and $\mathbf{Q} = (-5/2,-5/2,3/2)$ reflections stabilized in the respective $m = 1/4$ plateau phases are provided in (g-h). The kink highlighted by the red arrow in (i) matches the transition field between $m = 1/4$ SF and $m = 1/2$ DSF plateau phases. Data in (a,c-g) are taken from \cite{flavianMagneticPhaseDiagram2023a}.}
		\label{fig:Supp_DiffScans}
	\end{figure*}
	
	For the sake of completeness, in Fig.~\ref{fig:Supp_DiffScans} we provide a summary of the various magnetic superlattice peaks observed in \nbwo, including their field and temperature dependence (see Ref.~\cite{flavianMagneticPhaseDiagram2023a} for details). In zero field, a $\mathbf{q} = (0,0,1/3)$ propagation vector is realized (Fig.~\ref{fig:Supp_DiffScans}(a)), which turns incommensurate above a lock-in transition around $T^* \simeq 0.25$ K. Even above the ordering temperature $T_{\rm N}$, a broad diffuse scattering peak remains (Fig.~\ref{fig:Supp_DiffScans}(b)). As seen in Fig.~\ref{fig:Supp_DiffScans}(c-e), in an axial field $\mathbf{H \parallel c}$ the magnetic structure acquires an in-plane modulation, corresponding to a $\mathbf{q} = (1/3,1/3,1/3)$ propagation vector. Finally, for planar fields $\mathbf{H \parallel a^*}$ and $\mathbf{H \parallel a}$ respectively, a $\mathbf{q} = (0,1/2,1/2)$ and $\mathbf{q} = (1/2,1/2,1/2)$ order is realized (Fig.~\ref{fig:Supp_DiffScans}(f-i)). We point that the $\mathbf{H \parallel a}$ field dependence of the $\mathbf{Q} = (-5/2,-5/2,3/2)$ reflection shown in Fig.~\ref{fig:Supp_DiffScans}(i) exhibits a kink around 1.1 T, matching the transition field between the $m = 1/4$ and $m = 1/2$ plateau phases. 
	
	For the magnetic structures in applied fields, the presence of several independent orbits results in as many as 18 free parameters for a general field direction, precluding a magnetic Rietveld refinement based on symmetry analysis. Although our 1D spin-tube model does not make predictions on the {\it inter}-tube spin configurations, the combined diffraction data for the magnetized phases remain fully consistent with the {\it intra}-tube arrangements proposed in the main text. Due to the dominance of {\it intra}-tube exchanges $J_1, J_1' \gg J_\Delta, J_\nabla$ (see Sec.~\ref{Sec:3D_MF}), the 3D magnetic structures are likely constructed by stacking together different magnetic domains of the single-tube ground state configurations shown in Fig.~\ref{fig:Supp_SpinConfig}. For $\mathbf{H \parallel c}$, there are three magnetic domains, corresponding to $120^\circ$ rotations of the $\uparrow \uparrow \downarrow$ spiral state in Fig.~\ref{fig:Supp_SpinConfig}(b). As for the $\mathbf{H \parallel a, a^*}$ spin-flop structures (Fig.~\ref{fig:Supp_SpinConfig}(c,d)), a general field orientation selects two $180^\circ$ domains where only the "flopped" spins remain unchanged. In both cases, the predicted number of domain configurations matches the observed in-plane period of two/three unit cells, while also the experimental $c$-axis propagation vector component $q_c$ and fractional magnetization $m$ match the model predictions.

	\section{$\mathbf{H \parallel a}$ Presaturation Phase}
	
	As additional evidence of the $m = 1/2$ double spin-flop presaturation phase discussed in the main text, we carried out field-dependent dilatometry and torque measurements in $\mathbf{H \parallel a}$ configuration.
	
	For the torque experiment, we used a Faraday balance setup \cite{blosserFaradayMagnetometer2020} with a $0.8$ mg single crystal sample of \nbwo mounted on a flexible cantilever. Upon applying a field, there is a torque
	\begin{gather}
		\mathbf{\tau} = [\mathbf{m} \times \mathbf{H}] + [\mathbf{L} \times (\mathbf{m} \cdot \nabla) \mathbf{H}]
	\end{gather}
	acting on the magnetic moments $\mathbf{m}$ in the sample, where $\mathbf{L}$ is a vector connecting the fixed point on the cantilever to the sample position. The resulting cantilever deflection is translated into a change in capacitance, which is picked up by an Andeen-Hagerling 2550A bridge. 
	
	For the striction experiment, the change in sample length $\Delta L$ was measured with a minature capacitive dilatometer \cite{kuchlerNewDilatometer2023} in conjunction with the same AH 2550A bridge. The latter was operated at 1111 Hz to eliminate mechanical resonances that may limit the measurement precision. We measured the dilation along the $\mathbf{c}$-axis on a \nbwo single crystal of length $L_0 = 1.01$ mm, perpendicular to the field direction $\mathbf{H \parallel a}$. Both measurements are performed in a Quantum Design Physical Property Measurement System (PPMS) with 3He-4He dilution refrigerator insert at temperatures $T \lesssim 150$ mK.
	
\begin{figure}[tbp]
		\includegraphics[scale=1]{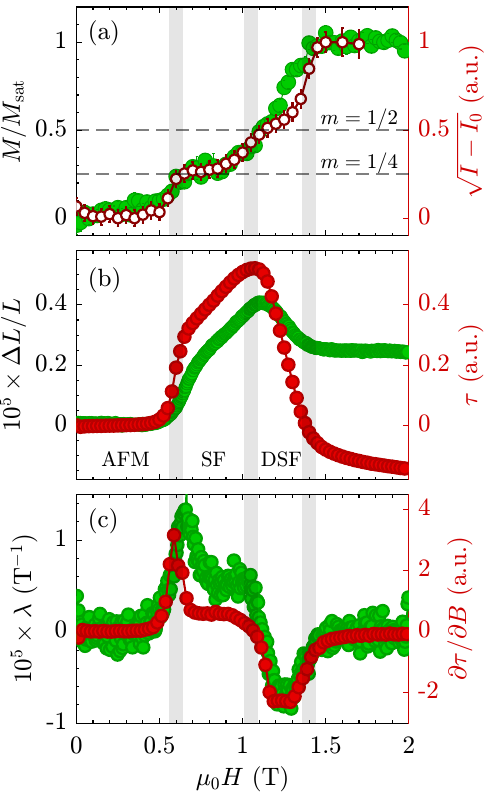}
		\caption{Field induced behavior of Nd$_3$BWO$_9$ in $\mathbf{H \parallel a}$ configuration at $T \lesssim 150$ mK. Grey vertical lines are guides to the eye representing the transition fields. (a) Magnetization curves obtained through neutron diffraction (red) and Faraday balance magnetometry (green). (b) Magnetic torque (red) and magneto-dilation along the $\mathbf{c}$-axis (green). (c) Torque derivative (red) and magnetostriction coefficient (green).}
		\label{fig:Supp_plateau}
	\end{figure}

	In Fig.~\ref{fig:Supp_plateau} we compare the torque and dilatometry data to the magnetization curve for $\mathbf{H \parallel a}$ (Fig.3(b) in the main text). Below the pseudospin saturation, both probes clearly exhibit three distinct field-regimes with constant slope, corresponding to the AFM, SF and DSF phases. The transition fields $h_{c1} = 0.60(5)$ T, $h_{c2} = 1.05(5)$ T and $h_{\rm sat} = 1.40(5)$ T are consistent among all techniques. The transitions themselves are even more pronounced in the numerically obtained magnetostriction coefficient $\lambda = \frac{1}{\mu_0 L} \frac{\partial \Delta L}{\partial H}$ and torque derivative, shown in Fig.~\ref{fig:Supp_plateau}(c). 
	
	\section{Low-Energy Neutron Spectroscopy}
		
\begin{figure}[tbp]
		\includegraphics[scale=1]{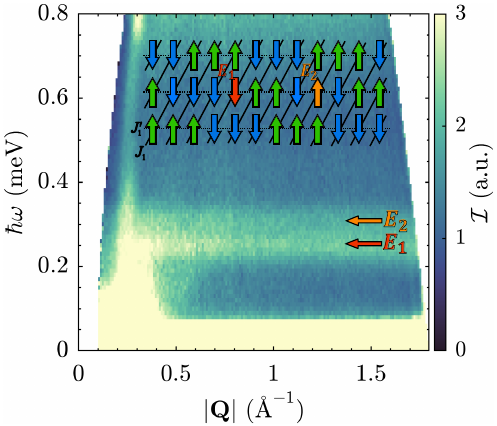}
		\caption{INS spectrum of \nbwo taken at $T = 40$ mK, showing two dispersionless modes. An inset depicts a sketch of the corresponding spin-flip excitations. The intense signal at $|\mathbf{Q}| < 0.4 \AA$ is spurious, resulting from an incomplete masking of the direct beam.}
		\label{fig:IN5}
	\end{figure}

	The low-energy spin excitation spectrum of \nbwo was investigated on the IN5 cold-neutron disk-chopper spectrometer at ILL. 340 single crystals were co-aligned in the $\mathbf{ac}$ scattering plane for a total sample mass of 1.58 g and installed in a dilution refrigerator at $T \lesssim 0.04$ K. Measurements were carried out at $E_i = 1.94$ meV (FWHM resolution $\sim 46$ $\mu$eV). As discussed in the main text, transitions between the dominant $\ket{j^z = \pm 9/2}$ states are forbidden by selection rules, meaning that only the small $\sim4\%$ deviation of the single-ion ground state wavefunctions from the ideal Ising limit contribute to the measured signal. Due to the very low scattering intensity, only a powder-average over all sample rotations and momentum transfers produced enough statistics for a quantitative analysis. This averaged  spectrum is shown in Fig.~\ref{fig:IN5}. One readily sees two dispersionless modes at $E_1 = 0.240(2)$~meV and $E_2 = 0.306(2)$ meV, whose intensity follows the square of the magnetic form factor. In our spin-tube model, these modes can be understood as single spin-flip excitations, where a spin at the edge (in the center) of an $\uparrow \uparrow \uparrow$ or $\downarrow \downarrow \downarrow$ spin cluster is flipped. These processes cost an energy $E_1 = J_1$ ($E_2 = J_1 - J_1'$) respectively (Fig.~\ref{fig:IN5}, inset). Based on the observed intensity ratio $\mathcal{I}_2 / \mathcal{I}_1 = 0.6(1)$, we can confidently assign $E_1 = J_1$ as the lower mode: For each magnetic unit cell there are 12(6) ways to create $E_1$($E_2$)-type spin flips, resulting in a twice stronger peak at $E_1$. This provides us with a direct measure of the dominant exchange parameters $J_1 \approx 0.240(2)$ meV and $J_1' \approx -0.28(2) J_1$, fairly close to the $J_1' \approx -0.35(5) J_1$ determined from the critical fields $h_c / h_{\rm sat} = 1 + J_1'/J_1$ delimiting the $m = 1/3$ magnetization plateau.
	The differences (of order $J_{\rm eff}$, see Sec.~\ref{Sec:3D_MF}) are likely caused by small in-plane couplings such as $J_\Delta, J_\nabla$, etc. For the INS modes we can calculate these effects directly based on the refined magnetic structure in zero field:
	\begin{gather}
		E_1 = J_1 \qquad\quad E_2 = J_1 - J_1' + J_\Delta + J_\nabla.
	\end{gather}
	As we can see $E_1 = J_1$ is exact in the classical limit, while the $E_2$-mode will be shifted linearly by other perturbing couplings. 
	
	\section{Numerical Modeling}
	
	In this section, we describe the various techniques employed to model the physics of \nbwo at the single spin-tube level and provide additional numerical simulation results. In total, we made use of four complementary approaches, all based on our classical Ising Hamiltonian for a single spin-tube. To understand the phase diagram, we performed brute force calculations of the classical ground state energy at $T = 0$. As a tool to model the thermodynamics at finite temperatures, we employed numerical transfer matrix calculations. The ordering susceptibility and spin correlations are simulated using classical Monte Carlo algorithm. Finally, to get an intuitive picture of the incommensurate correlations observed at elevated temperatures, we develop an analytical domain wall model. Except for the parameter sweeps to determine the phase diagram, all calculations were performed using $J_1'/J_1 = -0.35$ and $J_1 = 0.24$ meV.
	
	\subsection{$T = 0$ Phase Diagrams}

\begin{figure*}[tbp]
		\includegraphics[scale=1]{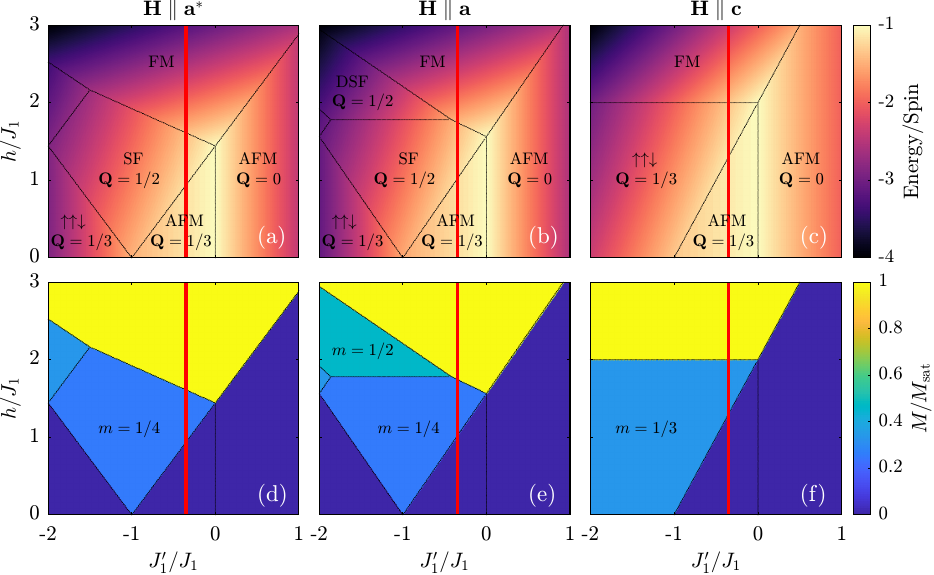}
		\caption{Classical phase diagram in our Ising spin-tube model for different exchange ratios $J_1'/J_1$ in an external magnetic field $h$ along three principal crystallographic axes. We plot the energy per spin (a-c) and the relative magnetization (d-f). The ratio $J_1' \simeq -0.35 J_1$ seen in experiment is shown as a red line. The phases are labeled as AFM (antiferromagnet), FM (ferromagnet), $\uparrow\uparrow\downarrow$ (up-up-down), SF (spin-flop) and DSF (double spin-flop), with fractional magnetization and $\mathbf{c}$-axis modulation labeled where relevant.}
		\label{fig:Supp_PD}
	\end{figure*}

\begin{figure*}[tbp]
		\includegraphics[scale=2]{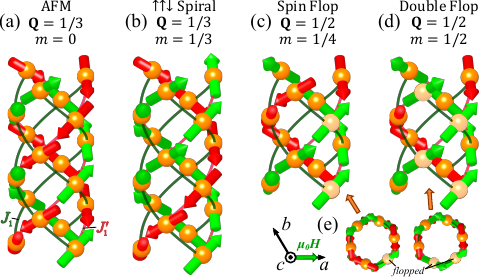}
		\caption{Magnetic structures of the single spin-tube model reproduced here for convenience, including (a) the AFM state (zero field), (b) the $\uparrow \uparrow \downarrow$ spiral phase ($\mathbf{H} \parallel \mathbf{c}$) and (c,d) the spin-flop (SF) and double spin-flop (DSF) phases ($\mathbf{H} \bot \mathbf{c}$). (e) Top-down view of the latter two structures.}
		\label{fig:Supp_SpinConfig}
	\end{figure*}

	Our one-dimensional Ising spin-tube model is simple enough that the ground state can be found through brute force calculations. We evaluate the energy for all possible spin configurations assuming a unit cell periodicity $\leq 4$, i.e. up to 24 spins per tube. In this way, we can construct a phase diagram at $T = 0$ versus exchange ratio $J_1'/J_1$ and magnetic field $h/J_1$. The phase boundaries can be extracted numerically by evaluating extrema in the second derivative of the ground state energy. The resulting phase diagrams for magnetic fields along the three principal crystallographic axes are shown in Fig.~\ref{fig:Supp_PD}(a-c). The magnetization can also be calculated as $M = \sum_i g_{zz} \mathbf{\hat{H}} \cdot \mathbf{\hat{z}}_i \hat{S}^z_i$ (see Fig.~\ref{fig:Supp_PD}(d-f)). For the sake of convenience, in Fig.~\ref{fig:Supp_SpinConfig} we reproduce the relevant spin-tube structures realized in \nbwo discussed in the main text.
	
\begin{figure}[tbp]
		\includegraphics[scale=1]{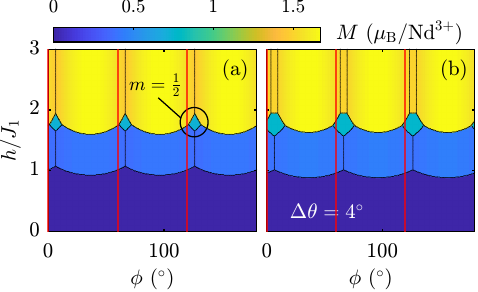}
		\caption{(a) Simulated magnetization versus field orientation using an exchange ratio $J_1'/J_1 = -0.35$ and an in-plane magnetic field $\mathbf{H \parallel ab}$, showing a small $m_z = 1/2$ magnetization plateau pocket centered at $h/J_1 \simeq 1.8$ and $\phi \simeq 6.5^\circ$ (modulo $60^\circ$). $\mathbf{H \parallel a}$ and equivalent configurations are marked as red vertical lines, showing that the plateau is narrowly avoided. (b) A small misalignment of the field towards the $\mathbf{c}$-axis tends to stabilize the $m_z = 1/2$ state. We show the same magnetization colorplot for a small offset $\Delta\theta = 4^\circ$, confirming the plateau is now fully stable.}
		\label{fig:Supp_angPD}
	\end{figure}

	Although the AFM $q = 1/3$ ground state, the $\uparrow\uparrow\downarrow$ plateau and the $m_z = 1/4$ spin-flop phases are all realized, the newly observed $m_z = 1/2$ double spin-flop phase seen along $\mathbf{H \parallel a}$ is barely avoided using our estimated exchange ratio $J_1' \simeq -0.35 J_1$. As seen in Fig.~\ref{fig:Supp_angPD}(a), it is stable only in a small pocket of in-plane field orientations for polar angles centered around $\phi = 6.5^\circ$ (modulo $60^\circ$), where four out of six spins have the same effective $g$-factor $g_{{\rm eff},i} = g_{zz} \mathbf{\hat{H}} \cdot \mathbf{\hat{z}}_i$ (the remaining two being completely decoupled from the field). In experiment, the crystal was aligned within $\sim 1^\circ$ in the $\mathbf{ac}$ horizontal scattering plane. But there is a large $\sim 5^\circ$ uncertainty in determining the orientation of the 1.8 T horizontal cryomagnet used to collect these data. In Fig.~\ref{fig:Supp_angPD}(b) we show the same magnetization map for a $\Delta \theta = 4^\circ$ misalignment from the $ab$-plane. Clearly, the $m_z = 1/2$ state is stabilized by even a small canting towards the $\mathbf{c}$-axis. The extreme sensitivity to the field orientation might explain why this plateau was missed in previous investigations \cite{flavianMagneticPhaseDiagram2023a, songMagneticFieldTuned2023}. Alternatively, it could be caused by a difference in temperature between the measurements - typically there is no hint of an incipient magnetization plateau above the ordering temperature.
	
	\subsection{Transfer Matrices \& Thermodynamics}
	
	\begin{figure*}[tbp]
		\includegraphics[scale=1]{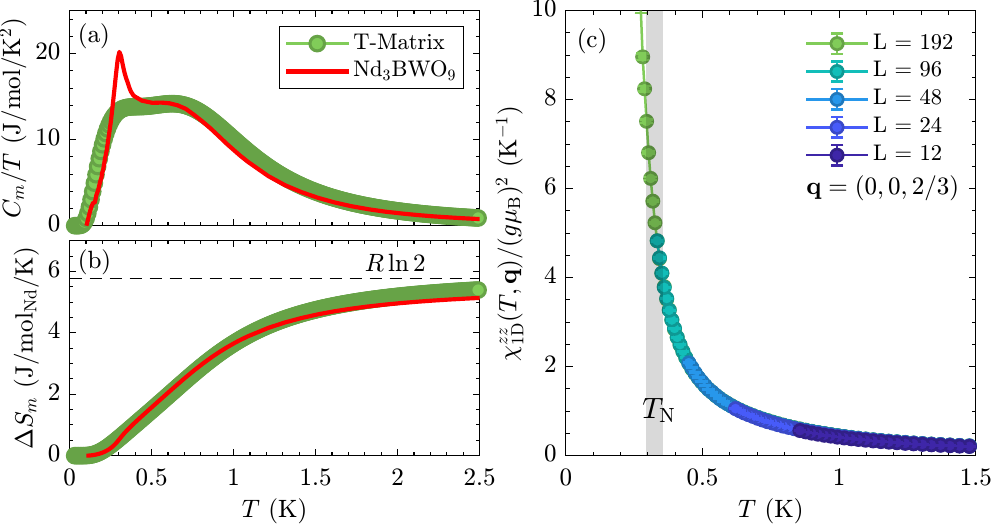}
		\caption{Zero-field numerical simulations of our 1D spin-tube model with $J_1=0.24$ meV and $J_1'/J_1 = -0.35$. (a,b) Transfer-matrix calculations of the magnetic specific heat $C_m$ and entropy change $\Delta S_m$ compared to experiment \cite{flavianMagneticPhaseDiagram2023a}. (c) Monte Carlo simulation of the ordering susceptibility $\chi_{\rm 1D}^{zz} (T,q)$ for various lattice sizes.}
		\label{fig:Supp_TMat_MC}
	\end{figure*}
	
	\begin{figure}[tbp]
		\includegraphics[scale=1]{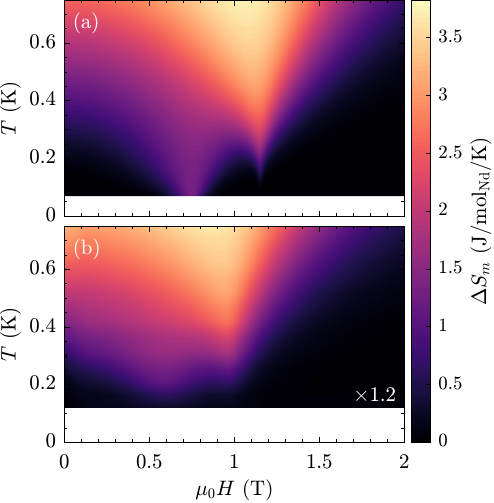}
		\caption{(a) Numerical transfer matrix calculation of the magnetic entropy change $\Delta S_m$ in an axial magnetic field $\mathbf{H \parallel c}$, visualized as a false colorplot. (b) Direct comparison to experimental entropy change in \nbwo for the same configuration (data taken from Ref. [10]). The data are scaled by $\times 1.2$ to obtain quantitative agreement with simulations.}
		\label{fig:Supp_TMat_dS}
	\end{figure}
	
	Since our 1D Ising spin-tube Hamiltonian can be decomposed into commuting, periodic clusters, the free energy of this model and its derivatives can be calculated exactly using the numerical transfer matrix technique \cite{kramersStatisticaFerromagnet1941}. Essentially, we break down our Hamiltonian into clusters of six spins, corresponding to the conventional unit cells of \nbwo. This allows us to write the transfer matrix
	\begin{gather}
		\mathbf{T}_{i,i+1} = \exp[-\beta V_{i,i}(J_1,J_1',h) -\beta V_{i,i+1}(J_1,J_1')],
	\end{gather}
	where $\beta = 1/k_B T$ is the inverse temperature, while $V_{i,i}$ and $V_{i,i+1}$ encode the interaction and Zeeman energies within a cluster and between adjacent clusters, respectively. The partition function for a tube made of $N$ unit cells can be expressed as $Z = {\rm Tr}(\mathbf{T}^N)$. In the thermodynamic limit $N \to \infty$, it is given by the leading eigenvalue $\lambda_0^N$. The free energy per site is then
	\begin{gather}
		f = -k_{\rm B} T \ln \lambda_0
	\end{gather}
	and can be evaluated exactly by numerically diagonalizing the $2^6 \times 2^6$ transfer matrix $\mathbf{T}$. Thermodynamic observables like the heat capacity or entropy are calculated directly as numerical derivatives of the free energy. The leading correlation length $\xi = 1/\ln(\lambda_0/|\lambda_1|)$ can be determined from the ratio of the two largest eigenvalues.

	In Fig.~\ref{fig:Supp_TMat_MC}(a,b) we present the calculated heat capacity $C_m$ and entropy change $\Delta S_m$ of our spin-tube model in zero magnetic field, compared directly to the experimental curves for \nbwo. Model and experiment show close agreement down to the ordering temperature $T_{\rm N}$. The broad peak in $C_p(T)$ around 0.8 K associated with the onset of short-range correlations is fully reproduced, while the sharp $\lambda$-anomaly at $T_{\rm N}$ is replaced by another hump in the spin-tube model.
	
	As discussed in the main text, the excellent agreement between model and experiment is maintained in a finite magnetic field $\mathbf{H \parallel c}$, seeing as the calculated heat capacity $C_p(H,T)$ semi-quantitatively captures all features across the phase diagram with no free parameters. In Fig.~\ref{fig:Supp_TMat_dS} we show that the same is true for the entropy change $\Delta S_m$, although here an overall scale factor $\times 1.2$ is necessary to match data and simulation. This is likely because the experimental curves (taken from \cite{flavianMagneticPhaseDiagram2023a}) are obtained indirectly by integrating the magnetocaloric field-sweeps, which may introduce some cumulative errors, e.g. through systematic deviations in the wire conductivity.
	
	Overall, the complete match of thermodynamic properties throughout the $H-T$ plane - and despite the lack of long-range ordering in one dimension - clearly emphasizes that our 1D spin-tube Hamiltonian captures the essential physics of \nbwo.
	
	\subsection{3D Effects \& Monte Carlo}
	\label{Sec:3D_MF}
	
	Our purely {\it one-dimensional} Ising spin-tube model cannot exhibit long-range order at finite temperatures. However, as in any real magnet, at low enough temperatures there will be additional 3D interactions between our 1D spin-tubes (e.g. the breathing Kagome couplings), which may induce a phase transition to an AFM ordered ground state. In mean-field theory, $T_{\rm N}$ can be estimated as $\chi_{\rm 1D} (T_{\rm N},q) = (g \mu_{\rm B})^2 / J_{\rm eff}$ from the staggered susceptibility at the ordering wavevector $q$, where $J_{\rm eff}$ corresponds to the non-frustrated component of the effective in-plane exchange. This should allow us to estimate the scale of {\it inter}-tube interactions.
	
	In principle, the transfer matrix method described above can be employed also to calculate susceptibilities or correlation functions. However, for the ordering susceptibility this is not practical: a staggered field with period three increases the size of the transfer matrix to $2^{18} \times 2^{18}$. Instead, we rely on classical Monte Carlo simulations to recover $\chi_{\rm 1D} (T,q)$. We model a single spin-tube, employing a simulated annealing procedure with $10^5$ Monte Carlo steps at each temperature. To exclude finite size-effects and ensure self-consistency, we simulate various lattice sizes up to $L = 192$ unit cells (i.e. $N \sim 10^4$ spins) and set a low-T cutoff where the 1D correlation length $\xi/c$ (obtained through the transfer matrix method) exceeds the number of sites in a spin-tube. The same procedure was used on a $L = 192$ tube to model the incommensurate spin correlations discussed in the main text. To ensure convergence, results were averaged over several simulation runs.
	
	The staggered susceptibility at $\mathbf{q} = (0,0,2/3)$ is depicted in Fig.~\ref{fig:Supp_TMat_MC}(c), diverging in a Curie-like manner. By invoking our mean-field condition $\chi_{\rm 1D} (T_{\rm N},q) = (g \mu_{\rm B})^2 / J_{\rm eff}$, we estimate $J_{\rm eff} \sim 0.2$ K, only about 7\% of the leading spin-tube exchange $J_1$. This strongly justifies our quasi-1D modeling approach, clearly implying that the {\it intra}-tube correlations dominate the physics of \nbwo.
			
	\subsection{Domain Wall Model}
	
	Below we present a short derivation of the Boltzmann domain wall model used to explain the incommensurate transition at $T^* \simeq 0.25$ K discussed in the main text. The lowest energy domain wall above the ground state exhibits a gap $\Delta E_{\rm DW} = (J_1 + J_1')/2 \approx 0.9$ K (see main text). Given that higher lying defects cost at least $\Delta E \gtrsim 2$ K to excite, these will be exponentially suppressed in the low temperature limit - i.e. only the lowest excitation is relevant close to $T^*$. Furthermore, in this dilute limit we can ignore interactions between domain walls. The partition function for a spin-tube composed of $N$ triangles ($N/2$ unit cells) can then be written as
	\begin{eqnarray}
		Z = \sum_{n = 0}^N g_n e^{-n \beta \Delta E_{\rm DW}}
	\end{eqnarray}
	where each set of triangles is either in the ground state configuration or forms a domain wall costing $\Delta E_{\rm DW}$. The degeneracy $g_n$ and Boltzmann weights $p_n$ are defined as
	\begin{eqnarray}
		g_n = \frac{N!}{(N-n)!n!}, \qquad p_n = \frac{1}{Z} g_n e^{-n \beta \Delta E_{\rm DW}}
	\end{eqnarray}
	and $\beta = 1/k_B T$. From this we can evaluate the expected number of defects $\langle n(T) \rangle = \sum_n n p_n$ at a given temperature $T$. The latter is directly related to the average periodicity of magnetic correlations
	\begin{eqnarray}
		\langle q (T) \rangle = \frac{1 + \langle n(T) \rangle / N}{3}.
	\end{eqnarray}
	We note that our domain wall model, being purely one-dimensional, does not predict a "real" transition, but a {\it crossover} from commensurate to incommensurate correlations. Nevertheless, the predicted crossover temperature nicely matches the experimental $T^* \simeq 0.25$ K with no free parameters, indicating that our model captures the essential physics.

	\bibliography{Nd3BWO9_bib}
	%\nocite{*}